\documentclass[11pt,english,onecolumn,draftcls]{IEEEtran}
\usepackage[T1]{fontenc}
\usepackage[latin9]{inputenc}
\usepackage{verbatim}
\usepackage{amsthm}
\usepackage{amsmath}
\usepackage{amssymb}
\usepackage{graphicx}

\makeatletter

\providecommand{\tabularnewline}{\\}

 \let\oldforeign@language\foreign@language
 \DeclareRobustCommand{\foreign@language}[1]{%
   \lowercase{\oldforeign@language{#1}}}
  \theoremstyle{remark}
  \newtheorem{rem}{\protect\remarkname}
  \theoremstyle{definition}
  \newtheorem{defn}{\protect\definitionname}
  \theoremstyle{plain}
  \newtheorem{thm}{\protect\theoremname}
  \theoremstyle{plain}
  \newtheorem{cor}{\protect\corollaryname}
  \theoremstyle{plain}
  \newtheorem{lem}{\protect\lemmaname}

\usepackage{cite}
\usepackage{times}

\newtheorem{assumption}{Assumption}

\makeatother

\usepackage{babel}
\providecommand{\corollaryname}{Corollary}
\providecommand{\definitionname}{Definition}
\providecommand{\lemmaname}{Lemma}
\providecommand{\remarkname}{Remark}
\providecommand{\theoremname}{Theorem}

\begin{document}

\title{Duality and Optimization for Generalized Multi-hop
MIMO Amplify-and-Forward Relay Networks with Linear Constraints}

\author{{\normalsize An Liu$^{1}$, Vincent K. N. Lau$^{1}$, Youjian Liu$^{2}$\\}$^{1}${\normalsize Department
of Electronic and Computer Engineering, Hong Kong University of Science
and Technology\\$^{2}$Department of Electrical and Computer Engineering,
University of Colorado at Boulder }%
\thanks{The work is funded by Huawei Technologies.%
}\vspace{-0.25in}
}
\maketitle
\begin{abstract}
We consider a generalized multi-hop MIMO amplify-and-forward (AF)
relay network with multiple sources/destinations and arbitrarily number
of relays. We establish two dualities and the corresponding \textit{dual
transformations} between such a network and its dual, respectively
under single network linear constraint and per-hop linear constraint.
The result is a generalization of the previous dualities under different
special cases and is proved using new techniques which reveal more
insight on the duality structure that can be exploited to optimize
MIMO precoders. A unified optimization framework is proposed to find
a stationary point for an important class of non-convex optimization
problems of AF relay networks based on a \textit{\normalsize local
Lagrange dual method}, where the \textit{primal algorithm} only finds
a stationary point for the inner loop problem of maximizing the Lagrangian
w.r.t. the primal variables. The input covariance matrices are shown
to satisfy a \textit{polite water-filling structure} at a stationary
point of the inner loop problem. The duality and polite water-filling
are exploited to design fast primal algorithms. Compared to the existing
algorithms, the proposed optimization framework with duality-based
primal algorithms can be used to solve more general problems with
lower computation cost.\end{abstract}
\begin{IEEEkeywords}
Multi-hop MIMO Networks, Amplify and Forward, Relay, Duality, MIMO
Precoder Optimization
\end{IEEEkeywords}

\section{Introduction}

The amplify and forward (AF) relay technique is useful in wireless
systems for cost-effective throughput enhancement and coverage extension.
It has attracted many research works recently. The optimization of
the MIMO (Multiple-input multiple-output) AF relay system has been
well studied (see the tutorial paper \cite{Sanguinetti_JSAC2012_AFsurvey}
and the reference there in). The optimal relay precoding matrix was
established in \cite{Hua_TWC2007_optAF} for a two-hop MIMO relay
system and the result was later extended to multi-hop MIMO relay system
in \cite{Yue_TWC10_NLrelay_DFE}, where the optimal source and relay
precoding matrices are shown to diagonalize the source-relay-destination
channel. It was shown in \cite{Rong_TSP11_AFQos} that this diagonalization
property also holds for power minimization of MIMO AF relay system
under QoS constraints. However, the power minimization problem is
still difficult to solve due to its non-convex nature\cite{Sanguinetti_TSP12_optAF}.
A low complexity algorithm is proposed in \cite{Sanguinetti_TSP12_optAF}
to closely approximate the solution of this non-convex problem for
the two-hop case. The source and relay power allocation problem of
the multicarrier two-hop AF relay system was studied in \cite{Zhang_TOC11_optAFOFDM}.
All of the above works focus on the case with a single relay at each
hop. In \cite{Fu_TSP11_AFmultiRelays}, the power minimization under
QoS constraints was addressed for a two-hop MIMO relay system with
multiple relays. There are also some works on the optimization of
multiuser MIMO AF relay systems. For example, the multiuser MIMO MAC
(multiaccess channel) AF relay networks are addressed in \cite{Phan_TWC09_PowAFMAC,Yu_TSP10_GWFRL},
while the MIMO BC (broadcast channel) AF relay networks are investigated
in \cite{Yu_TSP10_GWFRL}. Since the optimal structure of the source
and relay precoding matrices is still unknown for multi-user MIMO
AF relay networks, most of the algorithms in \cite{Phan_TWC09_PowAFMAC,Yu_TSP10_GWFRL}
are based on standard optimization methods such as Geometric programming
\cite{Phan_TWC09_PowAFMAC} and logarithmic barrier method \cite{Boyd_04Book_Convex_optimization}.
For MAC/BC AF relay networks with single-antenna source/destination
nodes, two dualities with respect to a total network power constraint
and per-hop power constraint have been established in \cite{Jafar_TIT07_Dualrelay}.
The results were later generalized to two-hop MIMO MAC/BC AF relay
networks \cite{Jafar_TCOM10_DualRelay} and multi-hop MIMO AF relay
systems \cite{Rong_TWC11_dualrelay}. In some cases, the duality can
be used to simplify the network optimization problem, e.g., some applications
of the duality can be found in \cite{Jafar_TIT09_NoissCorrelationAF,Rong_TWC11_dualrelay}.

In this paper, we consider a generalized multi-hop MIMO AF relay network
called the \textit{B-MAC AF relay network}, which has multiple sources,
multiple destinations and arbitrarily number of relays. The B-MAC
AF relay network topology includes almost all the existing AF relay
networks as special cases and the optimization of such general network
has not been addressed in the literature. We derive structural properties
of the optimization problem (\textit{duality} and \textit{polite water-filling
(PWF)}) and exploit the duality and PWF structure to design efficient
source and relay precoding algorithms for general B-MAC AF relay networks.
For the special case of the single-hop \textit{B-MAC interference
network} (B-MAC IFN), the duality and PWF results have been established
in \cite{Liu_IT10s_Duality_BMAC,Liu_10sTSP_MLC}. However, due to
the key difference in topology (single-hop versus multi-hop), there
are various technical challenges and we cannot extend the existing
results and algorithms trivially from \cite{Liu_IT10s_Duality_BMAC,Liu_10sTSP_MLC}.
Specifically, due to the multi-hop topology, the achievable rate is
a complicated function of the input covariance and relay precoding
matrices. This makes both the proof of dualities and the optimization
of B-MAC AF relay networks much more challenging than the previous
works with simpler network topology. Furthermore, the optimization
constraints in the multi-hop B-MAC AF relay network is more complicated
than that in B-MAC IFN. For example, the \textit{per-hop power constraint}
and \textit{per-relay power constraint} are two practically important
constraints that only occur in the multi-hop network topology.

We propose several new techniques to solve these challenges. The main
contributions are listed below.
\begin{itemize}
\item \textbf{Structural Properties (Duality) of MIMO AF Relay Network Optimization:}
We show that the achievable regions of a B-MAC AF relay network and
its dual are the same under\textit{ single network linear constraint}
(Type I duality) or \textit{per-hop linear constraint} (Type II duality).
We also derive explicit \textit{dual transformations} to calculate
the dual input covariance and relay precoding matrices. It is difficult
to use the brute force computation approach in \cite{Jafar_TIT07_Dualrelay,Jafar_TCOM10_DualRelay,Rong_TWC11_dualrelay}
to prove the duality for general B-MAC AF relay networks. We propose
two new techniques, namely the \textit{network equivalence} and the
\textit{network dual scaling} technique, to prove the Type I and Type
II dualities. The proof is not only simpler but also reveals more
insight on the duality structure. The duality and the network equivalence
results are useful for MIMO precoder optimization in B-MAC AF relay
networks. First, they are used in Section \ref{sub:PWF_Structure}
to derive the PWF structure of the optimal input covariance matrices,
which can be exploited to design efficient input covariance optimization
algorithm as illustrated in Section \ref{sub:Polite-water-filling-based}.
Second, the duality is used in Section \ref{sub:AlgMACBC} and \ref{sub:Algorithm-for-multi-hop}
to simplify the optimization of relay precoding matrices in BC AF
relay networks.
\item \textbf{Efficient MIMO Precoder Optimization Algorithm for AF Relay
Networks:} Based on the duality structure established above, we propose
efficient algorithms for MIMO precoder optimization in general MIMO
AF relay networks. Since the duality is only established for single
network / per-hop linear constraint, we cannot directly exploit the
duality structure to handle general power constraints such as per-relay
power constraint. To solve this problem, we propose a unified optimization
framework to find a stationary point for an important class of B-MAC
AF relay network optimization problems. In such a framework, we only
need to design a primal algorithm to find a stationary point for the
\textit{unconstrained} inner loop problem of maximizing the Lagrangian.
Using the network equivalence and duality results, we then show that
at a stationary point of the inner loop problem, the input covariances
have the PWF structure, which is a generalization of the PWF structure
in B-MAC IFN \cite{Liu_10sTSP_MLC} to the B-MAC AF Relay network.
Finally, we propose several efficient primal algorithms based on duality
and PWF structure for two-hop, three-hop, and multi-hop AF relay networks.
The proposed algorithms have significant advantages w.r.t. conventional
step-size based iterative algorithms (such as the gradient ascent
method \cite{Bertsekas_book99_NProgramming} and the logarithmic barrier
algorithms in \cite{Yu_TSP10_GWFRL}) in terms of complexity and convergence
speed. It can also be used to solve a more general class of problems
than the existing AF relay network optimization algorithms in \cite{Phan_TWC09_PowAFMAC,Yu_TSP10_GWFRL,Sanguinetti_JSAC2012_AFsurvey}.
\end{itemize}

The rest of the paper is organized as follows. Section \ref{sec:System Model}
describes the system model, the achievable scheme and the preliminary
results on rate duality for B-MAC IFN. Section \ref{sec:Duality-Results}
presents the main results on the two types of dualities for multi-hop
B-MAC AF relay networks. In Section \ref{sec:Applications-on-Network},
the duality is applied to study the non-convex optimization of B-MAC
AF Relay networks. A unified optimization framework together with
several primal algorithms are proposed to find a stationary point
for an important class of B-MAC AF relay network optimization problems.
The performance of the proposed algorithms is verified by simulation
in Section \ref{sec:Simulation-Results}. Section \ref{sec:Conlusion}
concludes.

\section{System Model and Preliminaries\label{sec:System Model}}

\subsection{B-MAC AF Relay Network\label{sub:B-MAC-AF-Relay}}

\begin{figure*}
\begin{centering}
\textsf{\includegraphics[clip,scale=0.65]{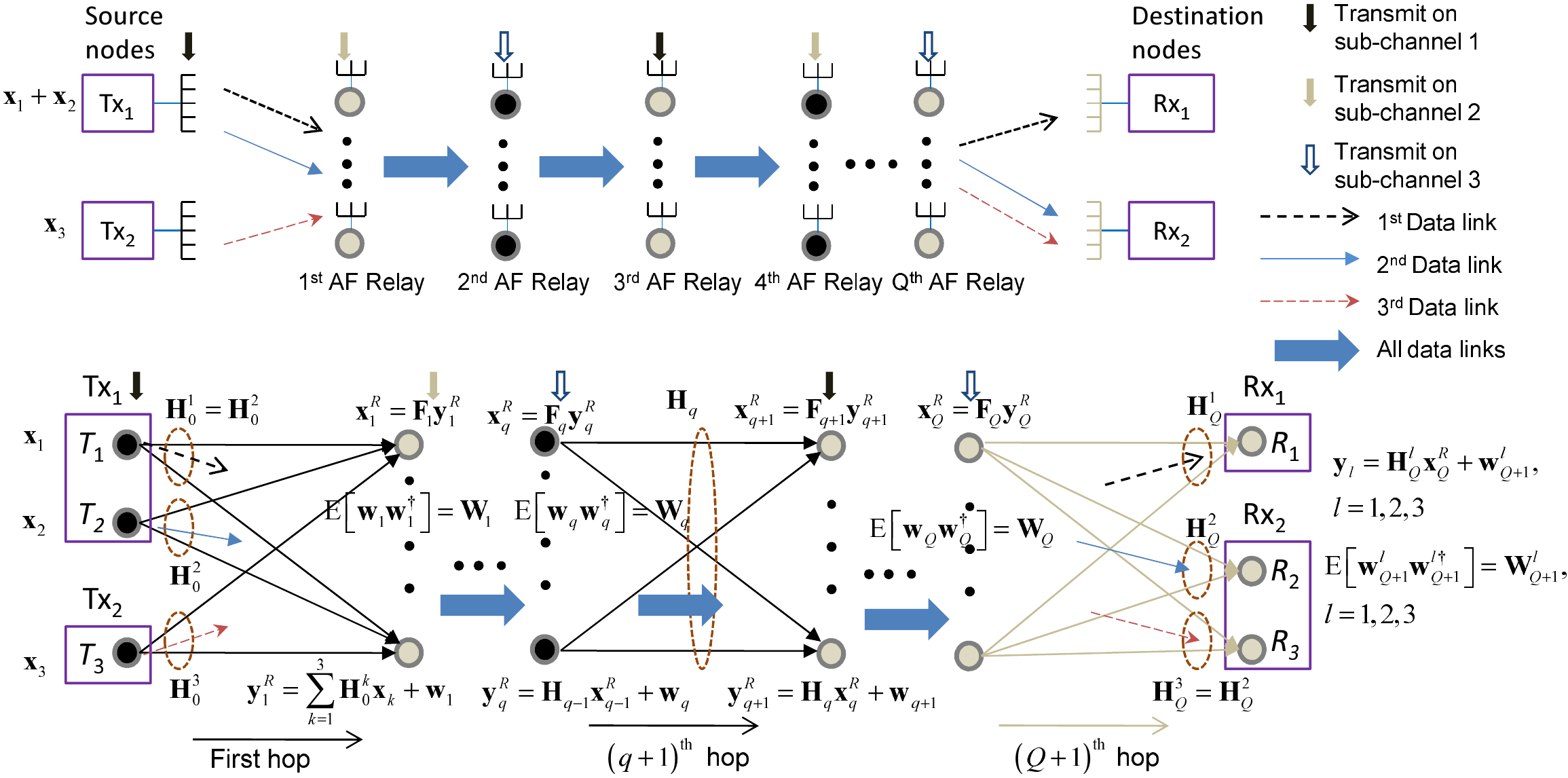}}
\par\end{centering}

\caption{\label{fig:B-MAC_Relay}An example of multi-hop B-MAC AF Relay Network
and its equivalent channel model }
\end{figure*}

A \textit{B-MAC AF relay network} is a general multi-hop AF relay
network with multiple sources (transmitters), multiple destinations
(receivers) and $Q$ relay clusters as illustrated in Fig. \ref{fig:B-MAC_Relay}.
Each relay cluster can have multiple relays. Each node can have multiple
antennas. We follow the widely used orthogonal assumption \cite{Jafar_TIT07_Dualrelay,Jafar_TCOM10_DualRelay,Rong_TWC11_dualrelay,Yu_TSP10_GWFRL}:

\begin{assumption}[Orthogonality among different hops]\label{asm:IFhop}Due
to propagation loss and proper channel reuse, the node in the $q^{\textrm{th}}$
cluster only receive the signals from the $\left(q-1\right)^{\textrm{th}}$
cluster, where the sources and destinations are considered as the
$0^{\textrm{th}}$ and $\left(Q+1\right)^{\textrm{th}}$ cluster respectively.\hfill \QED

\end{assumption}%

We consider $L$ data links between the sources and destinations.
Each source can have independent data for different destinations,
and each destination may want independent data from different sources.
Assume the %
sources are labeled by elements in $\mathcal{T}=\{\text{Tx}_{1},\text{Tx}_{2},\text{Tx}_{3},...\}$
and the destinations are labeled by elements in $\mathcal{R}=\{\text{Rx}_{1},\text{Rx}_{2},\text{Rx}_{3},...\}$.
Define $T_{l}$ ($R_{l}$) as the label of the source (destination)
of the $l^{\textrm{th}}$ data link.%
{} The numbers of antennas at $T_{l}$ and $R_{l}$ are $L_{T_{l}}$
and $L_{R_{l}}$ respectively. The total number of antennas at the
$q^{\textrm{th}}$ relay cluster is $L_{q}$. An example of B-MAC
AF relay network is illustrated in Fig. \ref{fig:B-MAC_Relay}. There
are 2 sources, 2 destinations, and 3 data links, of which $T_{1}=\text{Tx}_{1}$,
$T_{2}=\text{Tx}_{1}$, $T_{3}=\text{Tx}_{2}$, $R_{1}=\text{Rx}_{1}$,
$R_{2}=\text{Rx}_{2}$, and $R_{3}=\text{Rx}_{2}$. The figure also
illustrates how to properly allocate sub-channels to achieve the orthogonality
among different hops. In Fig. \ref{fig:B-MAC_Relay}, we assume that
due to path loss, the aggregate interference from the nodes which
are $2$ or more hops away from the receiver is negligible. Hence
we can avoid interference among the transmissions of different hops
by dividing the whole bandwidth into $3$ orthogonal sub-channels
and allocating different sub-channels to adjacent clusters.%

For convenience, we use link based notation for signals. Denote $\mathbf{H}_{0}^{l}\in\mathbb{C}^{L_{1}\times L_{T_{l}}}$
the channel matrix between source $T_{l}$ and the first relay cluster,
$\mathbf{H}_{q}\in\mathbb{C}^{L_{q+1}\times L_{q}},\:1\le q\le Q-1$
the overall channel matrix between the $q^{\textrm{th}}$ and $(q+1)^{\textrm{th}}$
relay cluster, and $\mathbf{H}_{Q}^{l}\in\mathbb{C}^{L_{R_{l}}\times L_{Q}}$
the channel matrix between the $Q^{\textrm{th}}$ relay cluster and
destination $R_{l}$. Let $\mathbf{x}_{k}\in\mathbb{C}^{L_{T_{k}}\times1}$
denote the transmit signal of link $k$ %
and assume it is a circularly symmetric complex Gaussian (CSCG) vector
with covariance $\mathbf{\Sigma}_{k}$. Let $\mathbf{w}_{q}\in\mathbb{C}^{L_{q}\times1}$
denote the overall CSCG noise vector at the $q^{\textrm{th}}$ relay
cluster with covariance $\mathbf{W}_{q}$. Then%
{} the composite received signal at the first relay cluster is $\mathbf{y}_{1}^{R}=\sum_{k=1}^{L}\mathbf{H}_{0}^{k}\mathbf{x}_{k}+\mathbf{w}_{1}$.
The $j^{\textrm{th}}$ relay in the $q^{\textrm{th}}$ cluster applies
linear transformation $\mathbf{F}_{q}\left(j\right)$ to the received
signal and the resulting composite transmit signal from the $q^{\textrm{th}}$
relay cluster is $\mathbf{x}_{q}^{R}=\mathbf{F}_{q}\mathbf{y}_{q}^{R}$,
where $\mathbf{F}_{q}=\textrm{Block Diag}\left[\mathbf{F}_{q}\left(1\right),...,\mathbf{F}_{q}\left(n_{q}\right)\right]\in\mathbb{C}^{L_{q}\times L_{q}}$
is the composite relay precoding matrix, $n_{q}$ is the number of
relays in the $q^{\textrm{th}}$ relay cluster, and $\mathbf{y}_{q}^{R}=\mathbf{H}_{q-1}\mathbf{x}_{q-1}^{R}+\mathbf{w}_{q}$
is the composite received signal at the $q^{\textrm{th}}$ relay cluster.
For convenience, define
\begin{eqnarray}
\mathbf{B}_{q,q^{'}} & = & \mathbf{F}_{q^{'}}\mathbf{H}_{q^{'}-1}\mathbf{F}_{q^{'}-1}...\mathbf{H}_{q}\mathbf{F}_{q},\:\forall q^{'}\geq q\label{eq:BDef}\\
\mathbf{\check{H}}_{l,k} & = & \mathbf{H}_{Q}^{l}\mathbf{B}_{1,Q}\mathbf{H}_{0}^{k},\:\forall k,l.\label{eq:EHlk}
\end{eqnarray}
Then%
{} using induction, it can be shown that the received signal at $R_{l}$
is

\begin{eqnarray}
\mathbf{y}_{l} & = & \sum_{k=1}^{L}\mathbf{\check{H}}_{l,k}\mathbf{x}_{k}+\sum_{q=1}^{Q}\mathbf{H}_{Q}^{l}\mathbf{B}_{q,Q}\mathbf{w}_{q}+\mathbf{w}_{Q+1}^{l},\label{eq:recvsignal}
\end{eqnarray}
where $\mathbf{w}_{Q+1}^{l}\in\mathbb{C}^{L_{R_{l}}\times1}$ is the
CSCG noise vector at $R_{l}$ with covariance $\mathbf{W}_{Q+1}^{l}$.
As an example, the equivalent channel model for the B-MAC AF relay
network in Fig. \ref{fig:B-MAC_Relay} is illustrated in the lower
sub-figure.

\begin{figure*}
\begin{centering}
\textsf{\includegraphics[clip,scale=0.7]{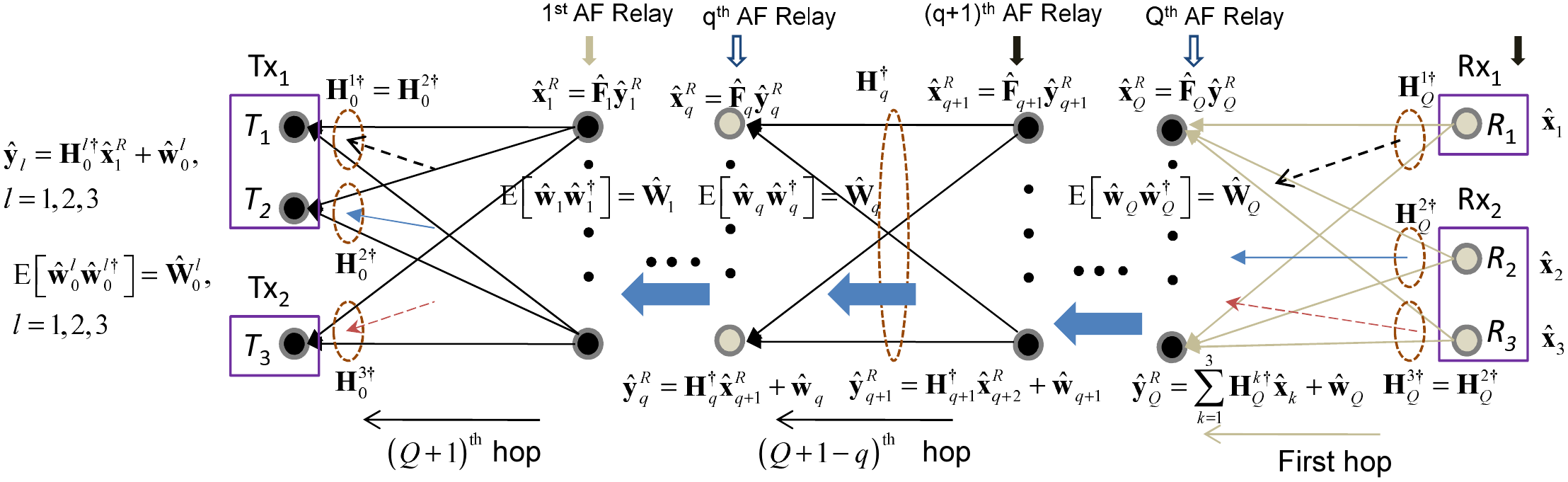}}
\par\end{centering}

\caption{\label{fig:B-MAC_Relay_Dual}Dual of the Multi-hop B-MAC AF Relay
Network in Fig. \ref{fig:B-MAC_Relay}}
\end{figure*}

\subsection{Transmit/Receive Scheme and Achievable Rate}

We consider a centralized optimization scheme in which a central node
computes all the input covariance and relay precoding matrices and
then transmits them to the source and relay nodes.

\begin{assumption}[Assumptions on the CSI]\label{asm:CSI}The central
node has the knowledge of the global channel state information (CSI)
$\mathbf{H}_{0}^{l}$'s $\mathbf{H}_{q}$'s and $\mathbf{H}_{Q}^{l}$'s.\hfill \QED

\end{assumption}

\begin{rem}
In two-hop AF relay systems, both the source-relay and relay-destination
channels can be estimated directly at the destination node by transmitting
pilot signals \cite{Hua_TSP11_AFchannelE}. However, acquiring global
CSI in multi-user multi-hop AF relay networks is difficult and requires
further study. In this paper, our main focus is to understand the
first order optimization problem and solution properties in multi-user
multi-hop AF networks. An interesting future work is to study efficient
decentralized solution which does not require global CSI knowledge.
The centralized solution in this paper provides the basis for studying
the decentralized solutions.%

\end{rem}

For fixed relay precoding matrix $\mathbf{F}_{q}$'s, the source nodes
and destinations see an equivalent channel in (\ref{eq:recvsignal}),
which is essentially a B-MAC IFN studied in \cite{Liu_IT10s_Duality_BMAC,Liu_10sTSP_Fairness_rate_polit_WF}.
Therefore, we can adopt the same set of transmit/receive schemes as
in \cite{Liu_IT10s_Duality_BMAC,Liu_10sTSP_Fairness_rate_polit_WF}.

\begin{assumption}[Assumptions on the transmit/receive scheme]\label{asm:TxRxscheme}$\:$
\begin{enumerate}
\item The transmit signal is CSCG.
\item Each signal is decoded by no more than one destination and can be
cancelled at this destination for the decoding of other signals.
\item The interference among the source-destination pairs can be specified
by some binary coupling matrix $\mathbf{\Phi}\in\{0,1\}^{L\times L}$.
That is, $\forall k,l$, after (possible) interference cancellation,
the interference from $\mathbf{x}_{k}$ to $\mathbf{x}_{l}$ is given
by $\phi_{l,k}\mathbf{\check{H}}_{l,k}\mathbf{x}_{k}$, where $\phi_{l,k}$
is the $(l,k)^{\textrm{th}}$ element of $\mathbf{\Phi}$. Because
$\phi_{l,k}\in\{0,1\}$, each interference is either completely cancelled
or treated as noise.\hfill \QED
\end{enumerate}
\end{assumption}

The set of allowed transmit/receive schemes includes most interference
management techniques as special cases, such as spatial interference
reduction through beamforming, dirty paper coding (DPC) \cite{Costa_IT83_Dirty_paper}
at sources and successive interference cancellation (SIC) at destinations
\cite{Liu_IT10s_Duality_BMAC}. For example, in the B-MAC AF relay
network in Fig. \ref{fig:B-MAC_Relay}, suppose DPC is used at $\textrm{Tx}_{1}$
with $\mathbf{x}_{1}$ encoded after $\mathbf{x}_{2}$, and SIC is
used at $\textrm{Rx}_{2}$ with $\mathbf{x}_{3}$ decoded after $\mathbf{x}_{2}$.
Then $\phi_{1,3}$, $\phi_{2,1}$, $\phi_{2,3}$ and $\phi_{3,1}$
are 1 and all other elements of $\mathbf{\Phi}$ are 0.

For given input covariance matrices $\mathbf{\Sigma}=\left(\mathbf{\Sigma}_{1},\mathbf{\Sigma}_{2},...,\mathbf{\Sigma}_{L}\right)$
and relay precoding matrices $\mathbf{F}=\left(\mathbf{F}_{1},\mathbf{F}_{2},...,\mathbf{F}_{Q}\right)$,
the covariance matrix of the transmit signal at the $q^{\textrm{th}}$
relay cluster is
\[
\mathbf{\Sigma}_{q}^{R}=\sum_{l=1}^{L}\mathbf{B}_{1,q}\mathbf{H}_{0}^{l}\mathbf{\Sigma}_{l}\mathbf{H}_{0}^{l\dagger}\mathbf{B}_{1,q}^{\dagger}+\sum_{q^{'}=1}^{q}\mathbf{B}_{q^{'},q}\mathbf{W}_{q^{'}}\mathbf{B}_{q^{'},q}^{\dagger}.
\]
The interference-plus-noise covariance at $R_{l}$ is
\begin{eqnarray}
\mathbf{\Omega}_{l} & = & \mathbf{W}_{l}^{'}+\sum_{k=1}^{L}\phi_{l,k}\mathbf{\check{H}}_{l,k}\mathbf{\Sigma}_{k}\mathbf{\check{H}}_{l,k}^{\dagger},\label{eq:whiteMG}
\end{eqnarray}
where $\phi_{l,l}=0$ and 
\begin{equation}
\mathbf{W}_{l}^{'}=\sum_{q=1}^{Q}\mathbf{H}_{Q}^{l}\mathbf{B}_{q,Q}\mathbf{W}_{q}\mathbf{B}_{q,Q}^{\dagger}\mathbf{H}_{Q}^{l\dagger}+\mathbf{W}_{Q+1}^{l}.\label{eq:Wlp}
\end{equation}
The achievable rate of the $l^{\textrm{th}}$ data link is \cite{Telatar_EuroTrans_1999_MIMOCapacity}
\begin{eqnarray}
\mathcal{I}_{l}^{\Phi}\left(\mathbf{\Sigma},\mathbf{F}\right) & = & \textrm{log}\left|\mathbf{I}+\check{\mathbf{H}}_{l,l}\mathbf{\Sigma}_{l}\mathbf{\check{H}}_{l,l}^{\dagger}\mathbf{\Omega}_{l}^{-1}\right|,\label{eq:linkkMIG}
\end{eqnarray}
on which the achievable regions in this paper are based.

We consider two types of linear constraints: Define the \textit{single
network linear constraint} as
\begin{equation}
{\textstyle \sum_{l=1}^{L}}\textrm{Tr}\left(\mathbf{\mathbf{\Sigma}}_{l}\mathbf{\hat{W}}_{0}^{l}\right)+{\textstyle \sum_{q=1}^{Q}}\textrm{Tr}\left(\mathbf{\mathbf{\Sigma}}_{q}^{R}\mathbf{\hat{W}}_{q}\right)\leq P_{T},\label{eq:DefSNLC}
\end{equation}
where $\mathbf{\hat{W}}_{0}^{l}$ and $\mathbf{\hat{W}}_{q}$ are
positive semidefinite; the\emph{ }\textit{per-hop linear constraint}
is defined as 
\begin{equation}
{\textstyle \sum_{l=1}^{L}}\textrm{Tr}\left(\mathbf{\mathbf{\Sigma}}_{l}\mathbf{\hat{W}}_{0}^{l}\right)\leq P_{0},\textrm{Tr}\left(\mathbf{\mathbf{\Sigma}}_{q}^{R}\mathbf{\hat{W}}_{q}\right)\leq P_{q},\:1\le q\le Q.\label{eq:DefILC}
\end{equation}
The above linear constraints are more general than the sum/individual
power constraints considered in \cite{Jafar_TIT07_Dualrelay,Jafar_TCOM10_DualRelay,Rong_TWC11_dualrelay}.
The per-hop linear constraint is more relevant to practical systems.
However, single network linear constraint is also important in practice
because it can be used to handle the more general multiple linear
constraints using Lagrange multiplier method as will be shown in Section
\ref{sec:Applications-on-Network}.

\subsection{The Dual of B-MAC AF Relay Network\label{sub:The-Dual-network}}
\begin{defn}
[Dual of a B-MAC AF relay network]\label{def:DualnetDef} The dual
of a B-MAC AF relay network is obtained by the following operations.
1) Reverse the transmission directions, i.e., in the dual network,
$R_{l}$'s become the sources, $T_{l}$'s become the destinations,
and the transmit signal of $R_{l}$ travels through the $Q^{\textrm{th}}$
relay cluster, the $\left(Q-1\right)^{\textrm{th}}$ relay cluster,
..., the $1^{\textrm{th}}$ relay cluster, and finally reaches the
destination $T_{l}.$ 2) Replace all the channel matrices by their
conjugate transposes, i.e., the channel matrix between source $R_{l}$
and the $Q^{\textrm{th}}$ relay cluster is $\mathbf{H}_{Q}^{l\dagger}$,
the channel matrix between the $\left(q+1\right)^{\textrm{th}}$ relay
cluster and the $q^{\textrm{th}}$ relay cluster is $\mathbf{H}_{q}^{\dagger},\: q=Q-1,...,1$,
and the channel matrix between the $1^{\textrm{th}}$ relay cluster
and destination $T_{l}$ is $\mathbf{H}_{0}^{l\dagger}$. 3) In the
dual network, the covariance of the noise at the $q^{\textrm{th}}$
relay cluster is set as $\mathbf{\hat{W}}_{q}$ and the covariance
of the noise at the destination $T_{l}$ is set as $\mathbf{\hat{W}}_{0}^{l}$,
where $\mathbf{\hat{W}}_{q}$'s and $\mathbf{\hat{W}}_{0}^{l}$'s
are the linear constraint matrices in (\ref{eq:DefSNLC}) of the original
network. 4) The coupling matrix for the dual network is set as the
transpose of that for the original network, i.e., $\mathbf{\Phi}^{T}$.%
\hfill \QED
\end{defn}

As an example, the dual of the B-MAC AF relay network in Fig. \ref{fig:B-MAC_Relay}
is illustrated in Fig. \ref{fig:B-MAC_Relay_Dual}.

We use the notation $\hat{}$ to denote the terms in the dual network.
Given input covariance matrices $\mathbf{\hat{\Sigma}}=\left(\mathbf{\hat{\Sigma}}_{1},\mathbf{\hat{\Sigma}}_{2},...,\mathbf{\hat{\Sigma}}_{L}\right)$
at the sources and relay precoding matrices $\mathbf{\hat{F}}=\left(\mathbf{\hat{F}}_{1},\mathbf{\hat{F}}_{2},...,\mathbf{\hat{F}}_{Q}\right)$
at the relays, the covariance matrix of the transmit signal at the
$q^{\textrm{th}}$ relay cluster is
\[
\mathbf{\hat{\Sigma}}_{q}^{R}=\sum_{l=1}^{L}\mathbf{\hat{B}}_{q,Q}^{\dagger}\mathbf{H}_{Q}^{l\dagger}\mathbf{\hat{\Sigma}}_{l}\mathbf{H}_{Q}^{l}\mathbf{\hat{B}}_{q,Q}+\sum_{q^{'}=q}^{Q}\mathbf{\hat{B}}_{q,q^{'}}^{\dagger}\mathbf{\hat{W}}_{q^{'}}\mathbf{\hat{B}}_{q,q^{'}},
\]
where 
\begin{equation}
\mathbf{\hat{B}}_{q,q^{'}}=\mathbf{\hat{F}}_{q^{'}}^{\dagger}\mathbf{H}_{q^{'}-1}\mathbf{\hat{F}}_{q^{'}-1}^{\dagger}...\mathbf{H}_{q}\mathbf{\hat{F}}_{q}^{\dagger},\:\forall q^{'}\geq q.\label{eq:Bhead}
\end{equation}
The interference-plus-noise covariance at $T_{l}$ is
\begin{equation}
\mathbf{\hat{\Omega}}_{l}=\mathbf{\hat{W}}_{0}^{l}+\sum_{q=1}^{Q}\mathbf{H}_{0}^{l\dagger}\mathbf{\hat{B}}_{1,q}^{\dagger}\mathbf{\hat{W}}_{q}\mathbf{\hat{B}}_{1,q}\mathbf{H}_{0}^{l}+\sum_{k=1}^{L}\phi_{k,l}\hat{\mathbf{H}}_{l,k}\mathbf{\hat{\Sigma}}_{k}\hat{\mathbf{H}}_{l,k}^{\dagger},\label{eq:whiteMG-1}
\end{equation}
where $\hat{\mathbf{H}}_{l,k}=\mathbf{H}_{0}^{l\dagger}\hat{\mathbf{B}}_{1,Q}^{\dagger}\mathbf{H}_{Q}^{k\dagger}$.
Hence, the achievable rate of the $l^{\textrm{th}}$ data link in
the dual network is \cite{Telatar_EuroTrans_1999_MIMOCapacity}
\begin{eqnarray}
\mathcal{\hat{I}}_{l}^{\Phi^{T}}\left(\mathbf{\hat{\Sigma}},\mathbf{\hat{F}}\right) & = & \textrm{log}\left|\mathbf{I}+\hat{\mathbf{H}}_{l,l}\mathbf{\hat{\Sigma}}_{l}\hat{\mathbf{H}}_{l,l}^{\dagger}\mathbf{\hat{\Omega}}_{l}^{-1}\right|.\label{eq:linkkMIG-1}
\end{eqnarray}
Similarly, in the dual network, the two types of linear constraints
are defined as 
\begin{equation}
{\textstyle \sum_{l=1}^{L}}\textrm{Tr}\left(\mathbf{\mathbf{\hat{\Sigma}}}_{l}\mathbf{W}_{Q+1}^{l}\right)+{\textstyle \sum_{q=1}^{Q}}\textrm{Tr}\left(\mathbf{\mathbf{\hat{\Sigma}}}_{q}^{R}\mathbf{W}_{q}\right)\leq P_{T},\label{eq:DefdualSNLC}
\end{equation}
\begin{equation}
{\textstyle \sum_{l=1}^{L}}\textrm{Tr}\left(\mathbf{\mathbf{\hat{\Sigma}}}_{l}\mathbf{W}_{Q+1}^{l}\right)\leq P_{Q},\textrm{Tr}\left(\mathbf{\mathbf{\hat{\Sigma}}}_{q}^{R}\mathbf{W}_{q}\right)\leq P_{q-1},\forall q.\label{eq:DefdualILC}
\end{equation}

\subsection{Rate Duality for One-hop B-MAC IFN\label{sec:Preliminary}}

The proof of the duality results in this paper relies on the rate
duality for B-MAC IFN \cite{Liu_IT10s_Duality_BMAC}, which is a special
case of the B-MAC AF relay network when $Q=0$. Let 
\begin{equation}
\left(\left[\mathbf{H}_{l,k}\right],{\textstyle \sum_{l=1}^{L}}\textrm{Tr}\left(\mathbf{\Sigma}_{l}\hat{\mathbf{W}}_{l}\right)\leq P_{T},\left[\mathbf{W}_{l}\right]\right),\label{eq:net-color-linear-constraint}
\end{equation}
denote a B-MAC IFN where the channel matrix between $T_{k}$ and $R_{l}$
is $\mathbf{H}_{l,k}$; the input covariance satisfies $\sum_{l=1}^{L}\textrm{Tr}\left(\mathbf{\Sigma}_{l}\hat{\mathbf{W}}_{l}\right)\leq P_{T}$;
the noise covariance at $R_{l}$ is $\mathbf{W}_{l}$. For fixed coupling
matrix $\mathbf{\Phi}$, the interference-plus-noise covariance and
the rate of link $l$ is given by 
\begin{eqnarray}
\mathbf{\Omega}_{l} & = & \mathbf{W}_{l}+\sum_{k=1}^{L}\phi_{l,k}\mathbf{H}_{l,k}\mathbf{\Sigma}_{k}\mathbf{H}_{l,k}^{\dagger},\label{eq:WhiteMRV-1}
\end{eqnarray}
and $\mathcal{I}_{l}^{\Phi}\left(\mathbf{\Sigma}\right)=\textrm{log}\left|\mathbf{I}+\mathbf{H}_{l,l}\mathbf{\Sigma}_{l}\mathbf{H}_{l,l}^{\dagger}\mathbf{\Omega}_{l}^{-1}\right|$
respectively. By definition, the dual network or \textit{reverse links}
is 
\begin{equation}
\left(\left[\mathbf{H}_{k,l}^{\dagger}\right],{\textstyle \sum_{l=1}^{L}}\textrm{Tr}\left(\hat{\mathbf{\Sigma}}_{l}\mathbf{W}_{l}\right)\leq P_{T},\left[\hat{\mathbf{W}}_{l}\right]\right).\label{eq:net-forward-color-dual}
\end{equation}
The interference-plus-noise covariance and the rate of reverse link
$l$ is given by 
\begin{eqnarray}
\hat{\mathbf{\Omega}}_{l} & = & \hat{\mathbf{W}}_{l}+\sum_{k=1}^{L}\phi_{k,l}\mathbf{H}_{k,l}^{\dagger}\mathbf{\hat{\mathbf{\Sigma}}}_{k}\mathbf{H}_{k,l},\label{eq:WhiteMRV}
\end{eqnarray}
and $\mathcal{\hat{I}}_{l}^{\mathbf{\Phi}^{T}}\left(\hat{\mathbf{\Sigma}}\right)=\textrm{log}\left|\mathbf{I}+\mathbf{H}_{l,l}^{\dagger}\hat{\mathbf{\Sigma}}_{l}\mathbf{H}_{l,l}\hat{\mathbf{\Omega}}_{l}^{-1}\right|$
respectively. 

The key of the rate duality is a \textit{covariance transformation}
\cite{Liu_IT10s_Duality_BMAC} defined as follows.
\begin{defn}
[Covariance transformation for B-MAC IFN]\label{def:The-covariance-transformation}
The covariance transformation of the input covariance $\mathbf{\Sigma}$
for a B-MAC IFN with parameters $\left\{ \left[\mathbf{H}_{l,k}\right],\left[\mathbf{W}_{l}\right],\left[\hat{\mathbf{W}}_{l}\right]\right\} $
as in (\ref{eq:net-color-linear-constraint}) is defined by 4 steps.

\textbf{\small Step }{\small 1: Decompose the signal of each link
$l$ to $M_{l}$ streams as 
\begin{equation}
\mathbf{\Sigma}_{l}={\textstyle \sum_{m=1}^{M_{l}}}p_{l,m}\mathbf{t}_{l,m}\mathbf{t}_{l,m}^{\dagger},l=1,...,L,\label{eq:Decomsig}
\end{equation}
where $\mathbf{t}_{l,m}\in\mathbb{C}^{L_{T_{l}}\times1}$ is a transmit
vector with $\left\Vert \mathbf{t}_{l,m}\right\Vert =1$ and $p_{l.m}$'s
are the transmit powers. }{\small \par}

\textbf{\small Step }{\small 2: $\forall l,m$, compute the MMSE-SIC
receive vector as \cite{Liu_IT10s_Duality_BMAC} 
\begin{align}
\mathbf{r}_{l,m} & =\alpha_{l,m}\left(\sum_{i=m+1}^{M_{l}}\mathbf{H}_{l,l}p_{l,i}\mathbf{t}_{l,i}\mathbf{t}_{l,i}^{\dagger}\mathbf{H}_{l,l}^{\dagger}+\mathbf{\Omega}_{l}\right)^{-1}\mathbf{H}_{l,l}\mathbf{t}_{l,m},\label{eq:MMSErev1G}
\end{align}
where $\alpha_{l,m}$ is chosen such that $\left\Vert \mathbf{r}_{l,m}\right\Vert =1$.
Calculate the SINR $\gamma_{l,m}$ as a function of $\left\{ p_{l,m},\mathbf{t}_{l,m},\mathbf{r}_{l,m}\right\} $
\cite{Liu_IT10s_Duality_BMAC}. }{\small \par}

\textbf{\small Step }{\small 3: In the reverse links, use $\left\{ \mathbf{r}_{l,m}\right\} $
as transmit vectors and $\left\{ \mathbf{t}_{l,m}\right\} $ as receive
vectors. Apply the SINR duality \cite{Rao_TOC07_netduality} to obtain
the reverse transmit powers $\mathbf{q}=\left[q_{1,1},...,q_{1,M_{1}},...,q_{L,1},...,q_{L,M_{L}}\right]^{T}$:
\begin{eqnarray}
\mathbf{q} & = & \left(\mathbf{D}^{-1}-\mathbf{\Psi}^{T}\right)^{-1}\mathbf{\hat{n}},\label{eq:qpower}
\end{eqnarray}
where $\mathbf{\Psi}\in\mathbb{R}_{+}^{\sum_{l}M_{l}\times\sum_{l}M_{l}}$
and the $\left(\sum_{i=1}^{l-1}M_{i}+m\right)^{\textrm{th}}$ row
and $\left(\sum_{i=1}^{k-1}M_{i}+n\right)^{\textrm{th}}$ column of
$\mathbf{\Psi}$ is 
\begin{align}
\mathbf{\Psi}_{l,m}^{k,n}= & \begin{cases}
0 & k=l\:\textrm{and}\: m\geq n,\\
\left|\mathbf{r}_{l,m}^{\dagger}\mathbf{H}_{l,l}\mathbf{t}_{l,n}\right|^{2} & k=l,\:\textrm{and}\: m<n,\\
\phi_{l,k}\left|\mathbf{r}_{l,m}^{\dagger}\mathbf{H}_{l,k}\mathbf{t}_{k,n}\right|^{2} & \textrm{otherwise},
\end{cases}\label{eq:faiG}
\end{align}
$\mathbf{D}\in\mathbb{R}_{+}^{\sum_{l}M_{l}\times\sum_{l}M_{l}}$
is a diagonal matrix with the $\left(\sum_{i=1}^{l-1}M_{i}+m\right)^{\text{th}}$
diagonal element given by }{\small \par}

{\small 
\begin{eqnarray}
\mathbf{D}_{\sum_{i=1}^{l-1}M_{i}+m,\sum_{i=1}^{l-1}M_{i}+m} & = & \gamma_{l,m}/\left|\mathbf{r}_{l,m}^{\dagger}\mathbf{H}_{l,l}\mathbf{t}_{l,m}\right|^{2},\label{eq:DG}
\end{eqnarray}
$\mathbf{\hat{n}}=\left[\hat{n}_{1,1},...,\hat{n}_{1,M_{1}},...,\hat{n}_{L,1},...,\hat{n}_{L,M_{L}}\right]^{T}$
with $\hat{n}_{l,m}=\mathbf{t}_{l,m}^{\dagger}\hat{\mathbf{W}}_{l}\mathbf{t}_{l,m}\:\forall l,m$.}{\small \par}

\textbf{\small Step }{\small 4: The }\emph{\small Covariance Transformation}{\small{}
from $\mathbf{\Sigma}$ to $\hat{\mathbf{\Sigma}}$ is}%
{\small 
\begin{align}
\hat{\mathbf{\Sigma}}_{l}={\textstyle \sum_{m=1}^{M_{l}}}q_{l,m}\mathbf{r}_{l,m}\mathbf{r}_{l,m}^{\dagger},l=1,...,L & .\label{eq:CovTrans}
\end{align}
}{\small \par}
\end{defn}

We restate the rate duality in \cite{Liu_IT10s_Duality_BMAC} as follows.
\begin{thm}
[Rate duality for B-MAC IFN]\label{thm:linear-color-dual} For any
$\mathbf{\Sigma}$ satisfying $\sum_{l=1}^{L}\textrm{Tr}\left(\mathbf{\Sigma}_{l}\hat{\mathbf{W}}_{l}\right)\leq P_{T}$
and achieving a rate point $\mathbf{r}$ in the network (\ref{eq:net-color-linear-constraint}),
its covariance transformation $\hat{\mathbf{\Sigma}}_{1:L}$ achieves
a rate point $\mathbf{\hat{r}}\geq\mathbf{r}$ in the dual network
(\ref{eq:net-forward-color-dual}) and satisfies $\sum_{l=1}^{L}\textrm{Tr}\left(\hat{\mathbf{\Sigma}}_{l}\mathbf{W}_{l}\right)=\sum_{l=1}^{L}\textrm{Tr}\left(\mathbf{\Sigma}_{l}\hat{\mathbf{W}}_{l}\right)$.
Thus, the achievable regions %
in the forward and reverse links are the same under a single linear
constraint.
\end{thm}

\section{Duality Results for Multi-hop B-MAC AF Relay Network\label{sec:Duality-Results}}

In this section, we establish two types of duality for multi-hop B-MAC
AF relay network respectively under single network linear constraint
and per-hop linear constraint.

\subsection{Type I Duality}

The type I duality can be established using the following\textit{
network equivalence} result. 
\begin{thm}
[Network Equivalence]\label{thm:Network-Equivalence}Fixing the
relay precoding matrices $\mathbf{F}$, the B-MAC AF relay network
in Section \ref{sub:B-MAC-AF-Relay} under constraint (\ref{eq:DefSNLC})
is equivalent to the following B-MAC IFN
\begin{equation}
\left(\left[\mathbf{\check{H}}_{l,k}\right],{\textstyle \sum_{l=1}^{L}}\textrm{Tr}\left(\mathbf{\Sigma}_{l}\hat{\mathbf{W}}_{l}^{'}\right)\leq P_{T}-P_{C},\left[\mathbf{W}_{l}^{'}\right]\right),\label{eq:EquiBmac}
\end{equation}
where $\mathbf{\check{H}}_{l,k},\:\forall k,l$ is defined in (\ref{eq:EHlk}),
$\mathbf{W}_{l}^{'}$ is defined in (\ref{eq:Wlp}), and
\begin{eqnarray}
\hat{\mathbf{W}}_{l}^{'} & = & {\textstyle \sum_{q=1}^{Q}}\mathbf{H}_{0}^{l\dagger}\mathbf{B}_{1,q}^{\dagger}\mathbf{\hat{W}}_{q}\mathbf{B}_{1,q}\mathbf{H}_{0}^{l}+\mathbf{\hat{W}}_{0}^{l},\label{eq:Whlp}\\
P_{C} & = & {\textstyle \sum_{q=1}^{Q}\sum_{q^{'}=q}^{Q}}b_{q,q^{'}},\nonumber \\
b_{q,q^{'}} & \triangleq & \textrm{Tr}\left(\mathbf{B}_{q,q^{'}}\mathbf{W}_{q}\mathbf{B}_{q,q^{'}}^{\dagger}\mathbf{\hat{W}}_{q^{'}}\right)\:\forall q^{'}\geq q.\label{eq:Defbq}
\end{eqnarray}
Similarly, fixing the dual relay precoding matrices $\mathbf{\hat{F}}=\left(\mathbf{F}_{1}^{\dagger},...,\mathbf{F}_{Q}^{\dagger}\right)$,
the dual B-MAC AF relay network in Definition \ref{def:DualnetDef}
under constraint (\ref{eq:DefdualSNLC}) is equivalent to the dual
network of (\ref{eq:EquiBmac}) given by
\begin{equation}
\left(\left[\mathbf{\check{H}}_{k,l}^{\dagger}\right],{\textstyle \sum_{l=1}^{L}}\textrm{Tr}\left(\hat{\mathbf{\Sigma}}_{l}\mathbf{W}_{l}^{'}\right)\leq P_{T}-P_{C},\left[\hat{\mathbf{W}}_{l}^{'}\right]\right).\label{eq:Dual_EquiBmac}
\end{equation}

\end{thm}

Please refer to Appendix \ref{sub:Proof-of-TheoremNE} for the proof.
Define the \textit{Type I dual transformation} as follows.
\begin{defn}
[Type I dual transformation]\label{def:TypeITf} For any input covariance
and relay precoding matrices $\mathbf{\Sigma},\ \mathbf{F}$, let
$\hat{\mathbf{\Sigma}}$ be the covariance transformation of $\mathbf{\Sigma}$
in Definition \ref{def:The-covariance-transformation} with parameters
$\left\{ \left[\mathbf{\check{H}}_{l,k}\right],\left[\mathbf{W}_{l}^{'}\right],\left[\hat{\mathbf{W}}_{l}^{'}\right]\right\} $.
Then $\hat{\mathbf{\Sigma}}$ and $\mathbf{\hat{F}}\triangleq\left(\mathbf{F}_{1}^{\dagger},...,\mathbf{F}_{Q}^{\dagger}\right)$
is called the \textit{Type I dual transformation} of $\mathbf{\Sigma},\ \mathbf{F}$.\hfill \QED
\end{defn}

Then it follows from Theorem \ref{thm:linear-color-dual} and \ref{thm:Network-Equivalence}
that the following theorem holds.
\begin{thm}
[Type I duality]\label{thm:typeIduality}For any input covariance
and relay precoding matrices $\mathbf{\Sigma},\ \mathbf{F}$ satisfying
$\sum_{l=1}^{L}\textrm{Tr}\left(\mathbf{\mathbf{\Sigma}}_{l}\mathbf{\hat{W}}_{0}^{l}\right)+\sum_{q=1}^{Q}\textrm{Tr}\left(\mathbf{\mathbf{\Sigma}}_{q}^{R}\mathbf{\hat{W}}_{q}\right)\triangleq P_{T}^{\textrm{tx}}\leq P_{T}$
and achieving a rate point $\mathbf{r}$ in a B-MAC AF relay network,
its Type I dual transformation $\hat{\mathbf{\Sigma}},\ \mathbf{\hat{F}}$
achieves a rate point $\mathbf{\hat{r}}\geq\mathbf{r}$ in the dual
network and satisfy $\sum_{l=1}^{L}\textrm{Tr}\left(\mathbf{\mathbf{\hat{\Sigma}}}_{l}\mathbf{W}_{Q+1}^{l}\right)+\sum_{q=1}^{Q}\textrm{Tr}\left(\mathbf{\mathbf{\hat{\Sigma}}}_{q}^{R}\mathbf{W}_{q}\right)=P_{T}^{\textrm{tx}}$.
Thus, the achievable regions in a B-MAC AF relay network under constraint
(\ref{eq:DefSNLC}) and that in the dual network under constraint
(\ref{eq:DefdualSNLC}) are the same.\hfill \QED
\end{thm}

\subsection{Type II Duality}

Type II duality can be proved from type I duality using the conception
of \textit{network dual scaling}. 
\begin{defn}
[Scaled dual network]A scaled dual network with scaling vector $\mathbf{d}=\left[d_{1},...,d_{Q},d_{Q+1}\right]^{T}$,
where $d_{q}>0,\: q=1,...,Q,$ and $d_{Q+1}=1$, is obtained by scaling
the covariance of the noise vectors in the dual network defined in
Section \ref{sub:The-Dual-network} as follows: the covariance of
the noise at the $q^{\textrm{th}}$ relay cluster is scaled to be
$d_{q+1}\mathbf{\hat{W}}_{q}$, and the covariance of the noise at
the destination $T_{l}$ is scaled to be $d_{1}\mathbf{\hat{W}}_{0}^{l}$.
Furthermore, the relay precoding matrices is given by  $\mathbf{\hat{F}}=\left(\mathbf{F}_{1}^{\dagger},...,\mathbf{F}_{Q}^{\dagger}\right)$.\hfill \QED 
\end{defn}

The following corollary follows immediately from Theorem \ref{thm:typeIduality}.
\begin{cor}
[Duality for scaled dual network]\label{cor:scalednet} For any
input covariance and relay precoding matrices $\mathbf{\Sigma},\ \mathbf{F}$
achieving a rate point $\mathbf{r}$ in a B-MAC AF relay network,
let $\hat{\mathbf{\Sigma}}\left(\mathbf{d}\right)$ be the covariance
transformation of $\mathbf{\Sigma}$ obtained by Definition \ref{def:The-covariance-transformation}
with parameters $\left\{ \left[\mathbf{\check{H}}_{l,k}\right],\left[\mathbf{W}_{l}^{'}\right],\left[\hat{\mathbf{W}}_{l}^{'}\left(\mathbf{d}\right)\right]\right\} $,
where 
\begin{eqnarray}
\hat{\mathbf{W}}_{l}^{'}\left(\mathbf{d}\right) & = & \sum_{q=1}^{Q}\mathbf{H}_{0}^{l\dagger}\mathbf{B}_{1,q}^{\dagger}d_{q+1}\mathbf{\hat{W}}_{q}\mathbf{B}_{1,q}\mathbf{H}_{0}^{l}+d_{1}\mathbf{\hat{W}}_{0}^{l}.\label{eq:Wheadd}
\end{eqnarray}
Then $\hat{\mathbf{\Sigma}}\left(\mathbf{d}\right)$  and $\mathbf{\hat{F}}=\left(\mathbf{F}_{1}^{\dagger},...,\mathbf{F}_{Q}^{\dagger}\right)$
achieves a rate point $\mathbf{\hat{r}}\geq\mathbf{r}$ in the scaled
dual network with scaling vector $\mathbf{d}$ and satisfy the constraint
\begin{equation}
\sum_{l=1}^{L}\textrm{Tr}\left(\mathbf{\mathbf{\hat{\Sigma}}}_{l}\left(\mathbf{d}\right)\mathbf{W}_{Q+1}^{l}\right)+\sum_{q=1}^{Q}\textrm{Tr}\left(\mathbf{\mathbf{\hat{\Sigma}}}_{q}^{R}\left(\mathbf{d}\right)\mathbf{W}_{q}\right)=\sum_{q=0}^{Q}d_{q+1}P_{q}^{\textrm{tx}},\label{eq:SNLCfixd}
\end{equation}
where 
\begin{eqnarray*}
\mathbf{\hat{\Sigma}}_{q}^{R}\left(\mathbf{d}\right) & = & \sum_{l=1}^{L}\mathbf{B}_{q,Q}^{\dagger}\mathbf{H}_{Q}^{l\dagger}\mathbf{\hat{\Sigma}}_{l}\left(\mathbf{d}\right)\mathbf{H}_{Q}^{l}\mathbf{B}_{q,Q}+\sum_{q^{'}=q}^{Q}d_{q^{'}+1}\mathbf{B}_{q,q^{'}}^{\dagger}\mathbf{\hat{W}}_{q^{'}}\mathbf{B}_{q,q^{'}},\\
P_{0}^{\textrm{tx}} & = & \sum_{l=1}^{L}\textrm{Tr}\left(\mathbf{\mathbf{\Sigma}}_{l}\mathbf{\hat{W}}_{0}^{l}\right),\: P_{q}^{\textrm{tx}}=\textrm{Tr}\left(\mathbf{\mathbf{\Sigma}}_{q}^{R}\mathbf{\hat{W}}_{q}\right),\: q=1,...,Q.
\end{eqnarray*}

\end{cor}

Another theorem is needed to prove the Type II duality. Obtain $\left\{ \mathbf{t}_{l,m}\right\} $,
$\left\{ \mathbf{r}_{l,m}\right\} $, $\mathbf{\Psi}$ and $\mathbf{D}$
using (\ref{eq:Decomsig}), (\ref{eq:MMSErev1G}), (\ref{eq:faiG})
and (\ref{eq:DG}) in Definition \ref{def:The-covariance-transformation}
with parameters $\left\{ \left[\mathbf{\check{H}}_{l,k}\right],\left[\mathbf{W}_{l}^{'}\right],\left[\hat{\mathbf{W}}_{l}^{'}\right]\right\} $.
Define
\begin{equation}
a_{q,q^{'}}=\mathbf{n}_{q}^{T}\left(\mathbf{D}^{-1}-\mathbf{\Psi}^{T}\right)^{-1}\hat{\mathbf{n}}_{q^{'}},\forall q,q^{'}\in\left\{ 0,1,...,Q\right\} ,\label{eq:Defaq}
\end{equation}
where $\mathbf{n}_{q}=\left[n_{1,1}^{q},...,n_{1,M_{1}}^{q},...,n_{L,1}^{q},...,n_{L,M_{L}}^{q}\right]^{T}$
and $\forall l,m$, 
\begin{eqnarray}
n_{l,m}^{q-1} & = & \mathbf{r}_{l,m}^{\dagger}\mathbf{H}_{Q}^{l}\mathbf{B}_{q,Q}\mathbf{W}_{q}\mathbf{B}_{q,Q}^{\dagger}\mathbf{H}_{Q}^{l\dagger}\mathbf{r}_{l,m},\ q=1,...,Q,\nonumber \\
n_{l,m}^{Q} & = & \mathbf{r}_{l,m}^{\dagger}\mathbf{W}_{Q+1}^{l}\mathbf{r}_{l,m},\label{eq:Defnq}
\end{eqnarray}
$\mathbf{\hat{n}}_{q}=\left[\hat{n}_{1,1}^{q},...,\hat{n}_{1,M_{1}}^{q},...,\hat{n}_{L,1}^{q},...,\hat{n}_{L,M_{L}}^{q}\right]^{T}$
and $\forall l,m$, 
\begin{eqnarray}
\hat{n}_{l,m}^{q} & = & \mathbf{t}_{l,m}^{\dagger}\mathbf{H}_{0}^{l\dagger}\mathbf{B}_{1,q}^{\dagger}\mathbf{\hat{W}}_{q}\mathbf{B}_{1,q}\mathbf{H}_{0}^{l}\mathbf{t}_{l,m},\: q=1,...,Q\nonumber \\
\hat{n}_{l,m}^{0} & = & \mathbf{t}_{l,m}^{\dagger}\mathbf{\hat{W}}_{0}^{l}\mathbf{t}_{l,m}.\label{eq:Defnqhead}
\end{eqnarray}
Define $\mathbf{A}\in\mathbb{R}_{+}^{\left(Q+1\right)\times\left(Q+1\right)}$
whose $(q,q^{'})^{\textrm{th}}$ element is 
\begin{equation}
\begin{cases}
\left(a_{q-1,q^{'}-1}+b_{q,q^{'}-1}\right)/P_{q-1}^{\textrm{tx}},\: & \textrm{if}\: q^{'}>q\\
a_{q-1,q^{'}-1}/P_{q-1}^{\textrm{tx}} & \textrm{otherwise}
\end{cases}.\label{eq:DefA}
\end{equation}

\begin{thm}
[Network Dual Scaling]\label{thm:Network-Dual-Scaling}Consider
the following eigensystem
\begin{equation}
\mathbf{A}\mathbf{\tilde{d}}^{'}=\lambda_{\textrm{max}}\mathbf{\tilde{d}}^{'},\label{eq:Keyeigensys}
\end{equation}
where $\mathbf{\tilde{d}}^{'}$ is the dominant eigenvector and $\lambda_{\textrm{max}}$
is the maximum eigenvalue of $\mathbf{A}$. Then $\lambda_{\textrm{max}}$
and $\mathbf{\tilde{d}}^{'}$ must be strictly positive. Let $\tilde{\mathbf{d}}=\left[\tilde{d}_{1},...,\tilde{d}_{Q},1\right]$
be the dominant eigenvector of $\mathbf{A}$ with the last component
scaled to one. Then in the scaled dual network with scaling vector
$\tilde{\mathbf{d}}$, $\hat{\mathbf{\Sigma}}\left(\tilde{\mathbf{d}}\right)$
defined in Corollary \ref{cor:scalednet} and $\mathbf{\hat{F}}=\left(\mathbf{F}_{1}^{\dagger},...,\mathbf{F}_{Q}^{\dagger}\right)$
satisfies
\begin{eqnarray}
{\textstyle \sum_{l=1}^{L}}\textrm{Tr}\left(\mathbf{\mathbf{\hat{\Sigma}}}_{l}\left(\tilde{\mathbf{d}}\right)\mathbf{W}_{Q+1}^{l}\right)=P_{Q}^{\textrm{tx}}\leq P_{Q},\label{eq:IndLconsdualtx}\\
\textrm{Tr}\left(\mathbf{\mathbf{\hat{\Sigma}}}_{q}^{R}\left(\tilde{\mathbf{d}}\right)\mathbf{W}_{q}\right)=d_{q}P_{q-1}^{\textrm{tx}}\leq d_{q}P_{q-1},\: q=1,...Q.\nonumber 
\end{eqnarray}

\end{thm}

The proof is given in Appendix \ref{sub:Proof-of-Theorem_NDS}. Define
the \textit{Type II dual transformation} as follows.
\begin{defn}
[Type II dual transformation]For any input covariance and relay
precoding matrices $\mathbf{\Sigma},\mathbf{F}$, let $\hat{\mathbf{\Sigma}}\triangleq\hat{\mathbf{\Sigma}}\left(\tilde{\mathbf{d}}\right)$
be the covariance transformation of $\mathbf{\Sigma}$ in Definition
\ref{def:The-covariance-transformation} with parameters $\left\{ \left[\mathbf{\check{H}}_{l,k}\right],\left[\mathbf{W}_{l}^{'}\right],\left[\hat{\mathbf{W}}_{l}^{'}\left(\mathbf{\tilde{d}}\right)\right]\right\} $,
where $\tilde{\mathbf{d}}=\left[\tilde{d}_{1},...,\tilde{d}_{Q},1\right]$
is the dominant eigenvector of the eigensystem in (\ref{eq:Keyeigensys}).
Then $\hat{\mathbf{\Sigma}}$ and the dual relay precoding matrices
$\mathbf{\hat{F}}\triangleq\mathbf{\hat{F}}\left(\tilde{\mathbf{d}}\right)=\left(\tilde{c}_{1}\mathbf{F}_{1}^{\dagger},...,\tilde{c}_{Q}\mathbf{F}_{Q}^{\dagger}\right)$,
where $\tilde{c}_{Q}=1/\sqrt{\tilde{d}_{Q}}$, $\tilde{c}_{q}=\sqrt{\tilde{d}_{q+1}/\tilde{d}_{q}},\: q=1,...,Q-1$,
is called the \textit{Type II dual transformation} of $\mathbf{\Sigma},\mathbf{F}$.\hfill \QED
\end{defn}

For the same input covariance $\hat{\mathbf{\Sigma}}\triangleq\hat{\mathbf{\Sigma}}\left(\tilde{\mathbf{d}}\right)$,
the rate point in the scaled dual network with scaling vector $\tilde{\mathbf{d}}$
and relay precoding matrices $\left(\mathbf{F}_{1}^{\dagger},...,\mathbf{F}_{Q}^{\dagger}\right)$
is equal to that in the original dual network with relay precoding
matrices $\mathbf{\hat{F}}\triangleq\mathbf{\hat{F}}\left(\tilde{\mathbf{d}}\right)$.
Furthermore, it follows from (\ref{eq:IndLconsdualtx}) that the Type
II dual transformation $\hat{\mathbf{\Sigma}},\mathbf{\hat{F}}$ satisfies
(\ref{eq:DefdualILC}). Combining the above and Corollary \ref{cor:scalednet},
the following theorem is proved.
\begin{thm}
[Type II duality]\label{thm:typeIIduality}For any input covariance
and relay precoding matrices $\mathbf{\Sigma},\mathbf{F}$ satisfying
$\sum_{l=1}^{L}\textrm{Tr}\left(\mathbf{\mathbf{\Sigma}}_{l}\mathbf{\hat{W}}_{0}^{l}\right)\triangleq P_{0}^{\textrm{tx}}\leq P_{0},$
$\textrm{Tr}\left(\mathbf{\mathbf{\Sigma}}_{q}^{R}\mathbf{\hat{W}}_{q}\right)\triangleq P_{q}^{\textrm{tx}}\leq P_{q},\:\forall q$,
and achieving a rate point $\mathbf{r}$ in the B-MAC AF relay network
defined in Section \ref{sub:B-MAC-AF-Relay}, its Type II dual transformation
$\hat{\mathbf{\Sigma}},\mathbf{\hat{F}}$ achieves a rate point $\mathbf{\hat{r}}\geq\mathbf{r}$
in the dual network and satisfy $\sum_{l=1}^{L}\textrm{Tr}\left(\mathbf{\mathbf{\hat{\Sigma}}}_{l}\mathbf{W}_{Q+1}^{l}\right)=P_{Q}^{\textrm{tx}},$
$\textrm{Tr}\left(\mathbf{\mathbf{\hat{\Sigma}}}_{q}^{R}\mathbf{W}_{q}\right)=P_{q-1}^{\textrm{tx}},\:\forall q$.
{} Thus, the achievable regions in a B-MAC AF relay network under constraint
(\ref{eq:DefILC}) and that in the dual network under constraint (\ref{eq:DefdualILC})
are the same.\end{thm}
\begin{rem}
[Generality of Type I/II duality]\label{rem:Generality-of-TypeI_II}The
previous duality results for various special cases are compared to
illustrate the generality of the Type I/II duality. In \cite{Jafar_TIT07_Dualrelay}
and \cite{Jafar_TCOM10_DualRelay}, the duality was established respectively
for multi-hop MAC/BC AF relay networks with single-antenna source/destination
nodes and two-hop MIMO MAC/BC AF relay networks. However, the approach
in \cite{Jafar_TIT07_Dualrelay,Jafar_TCOM10_DualRelay} cannot be
easily extended to the general B-MAC AF relay network.\textbf{ }The
duality for multi-hop MIMO AF relay system was established in \cite{Rong_TWC11_dualrelay}.
Although the proof can be extended to multi-user AF relay networks
by using block diagonal precoding / receiving matrices and the notion
of independent streams, the duality in {[}3{]} cannot cover the duality
in this paper as elaborated below.

1) The B-MAC AF relay network is not captured by the system model
in \cite{Rong_TWC11_dualrelay}. In \cite{Rong_TWC11_dualrelay},
there are $N_{b}$ independent data streams in one transmission. Similar
to the coupling matrix defined in Assumption \ref{asm:TxRxscheme},
we can use a binary \textit{inter-stream coupling matrix} $\mathbf{\Phi}^{s}\in\{0,1\}^{N_{b}\times N_{b}}$
to specify the interference among the data streams. In \cite{Rong_TWC11_dualrelay},
$\mathbf{\Phi}^{s}$ can only have two forms: a) if linear transceivers
are used at the source and destination nodes, all the diagonal elements
of $\mathbf{\Phi}^{s}$ are zero, and all the off-diagonal elements
of $\mathbf{\Phi}^{s}$ are one; b) if non-linear transceivers are
used, $\mathbf{\Phi}^{s}$ is a triangular matrix. However, for general
B-MAC AF relay networks, if we decompose the equivalent MIMO links
into independent data streams, the inter-stream coupling matrix $\mathbf{\Phi}^{s}$
can be any binary matrix with zero diagonal elements.

2) Explicit dual transformations are not part of \cite{Rong_TWC11_dualrelay}
for the multi-user case with general inter-stream coupling matrix
$\mathbf{\Phi}^{s}$. The dual transformation between the DPC-based
and SIC-based MIMO AF relay systems in \cite{Rong_TWC11_dualrelay}
cannot be generalized to B-MAC AF relay networks because the matrix
``$\mathbf{\Phi}$'' in (14) of \cite{Rong_TWC11_dualrelay} is
no longer upper-triangle. 

Furthermore, the above special cases consider power constraints only.
By using the techniques of \textit{network equivalence} and \textit{network
dual scaling}, we are able to establish the dualities and explicit
dual transformations for B-MAC AF relay networks under more general
linear constraints. The duality proof based on these new techniques
is not only simpler but also reveals more insight on the duality structure
that can be exploited to design MIMO precoder optimization algorithms
as shown in the next section.
\end{rem}

\section{Applications in Network Optimization Problems\label{sec:Applications-on-Network}}

We propose an unified optimization framework to find a stationary
point for a class of AF relay network optimization problems based
on the local Lagrange dual method (LDM) \cite{Liu_11sTSP_MLC_localLDM},
where the \textit{primal algorithm} only finds a stationary point
for the inner loop problem of maximizing the Lagrangian. The duality
is first used to characterize the PWF structure of the input covariance
matrices at a stationary point. Then the duality and PWF are exploited
to design efficient primal algorithm for general \emph{B-MAC} AF relay
networks and structured primal algorithms for \emph{BC} AF Relay network.

\subsection{Unified Optimization Framework based on Local LDM}

A general optimization problem in a B-MAC AF Relay network can be
expressed as
\begin{equation}
\textrm{max}\: f\left(\mathbf{\Sigma},\mathbf{F}\right),\:\textrm{s.t.}\: g_{n}\left(\mathbf{\Sigma},\mathbf{F}\right)\geq0,\: n=1,...,N,\:\textrm{and}\:\mathbf{\Sigma}_{l}\succeq0,\: l=1,...,L,\label{eq:GeneralPF}
\end{equation}
where $f\left(\mathbf{\Sigma},\mathbf{F}\right)$ and $g_{n}\left(\mathbf{\Sigma},\mathbf{F}\right),\:\forall n$
are real valued functions of $\mathbf{\Sigma},\mathbf{F}$. In this
paper, we consider a class of problems whose Lagrangian can be written
in the following form
\begin{eqnarray}
L\left(\mathbf{\Sigma},\mathbf{F},\mathbf{\lambda}\right) & \triangleq & f\left(\mathbf{\Sigma},\mathbf{F}\right)+\sum_{n=1}^{N}\lambda_{n}g_{n}\left(\mathbf{\Sigma},\mathbf{F}\right)\nonumber \\
 & = & \sum_{l=1}^{L}w_{l}\mathcal{I}_{l}^{\Phi}\left(\mathbf{\Sigma},\mathbf{F}\right)-\sum_{l=1}^{L}\textrm{Tr}\left(\mathbf{\mathbf{\Sigma}}_{l}\mathbf{\hat{W}}_{0}^{l}\right)-\sum_{q=1}^{Q}\textrm{Tr}\left(\mathbf{\mathbf{\Sigma}}_{q}^{R}\mathbf{\hat{W}}_{q}\right),\:\mathbf{\mathbf{\Sigma}}_{l}\succeq0,\:\forall l,\label{eq:Lag}
\end{eqnarray}
where $\mathbf{\lambda}=\left[\lambda_{n}\geq0\right]_{n=1,...,N}$
are Lagrange multipliers, $\mathbf{\hat{W}}_{0}^{l},\:\forall l$
are positive semidefinite, $\mathbf{\hat{W}}_{q}\:\forall q$ are
positive definite, and $w_{l}\geq0,\:\forall l$. The Lagrangian of
many important optimization problems can be written in the form of
(\ref{eq:Lag}). Two typical problems are listed below.

P1: Maximize the weighted sum-rate under multiple linear constraints:
\begin{eqnarray}
 & \textrm{max} & \sum_{l=1}^{L}\mu_{l}\mathcal{I}_{l}^{\Phi}\left(\mathbf{\Sigma},\mathbf{F}\right)\label{eq:P1}\\
 & \text{s.t.} & \sum_{l=1}^{L}\textrm{Tr}\left(\mathbf{\Sigma}_{l}\mathbf{\hat{W}}_{n}^{l}\right)+\sum_{q=1}^{Q}\textrm{Tr}\left(\mathbf{\mathbf{\Sigma}}_{q}^{R}\mathbf{\hat{W}}_{n,q}\right)\leq\delta_{n},\ n=1,...,N,\:\textrm{and}\:\mathbf{\Sigma}_{l}\succeq0,\ \forall l,\nonumber 
\end{eqnarray}
where $\mu_{l}\geq0,\:\forall l$, $\mathbf{\hat{W}}_{n}^{l}$'s and
$\mathbf{\hat{W}}_{n,q}$'s are positive semidefinite and $\delta_{n}\geq0,\ \forall n$. 

P2: Minimize a single linear cost under individual rate constraints:
\begin{eqnarray}
 & \textrm{min} & \sum_{l=1}^{L}\textrm{Tr}\left(\mathbf{\mathbf{\Sigma}}_{l}\mathbf{\hat{W}}_{0}^{l}\right)+\sum_{q=1}^{Q}\textrm{Tr}\left(\mathbf{\mathbf{\Sigma}}_{q}^{R}\mathbf{\hat{W}}_{q}\right)\label{eq:P2}\\
 & \text{s.t.} & \mathcal{I}_{l}^{\Phi}\left(\mathbf{\Sigma},\mathbf{F}\right)>\mathcal{I}_{l}^{0},\ l=1,...,L,\:\textrm{and}\:\mathbf{\Sigma}_{l}\succeq0,\ \forall l,\nonumber 
\end{eqnarray}
where $\mathcal{I}_{l}^{0}\geq0$ is the rate constraint for the $l^{\textrm{th}}$
data link.
\begin{rem}
[Per-relay power constraint]The multiple linear constraints in P1
include the per-relay power constraint as special cases. Let $n_{q}$
denote the number of relays in the $q^{\textrm{th}}$ relay cluster,
and let $L_{q,j}$ denote the number of antennas at the $j^{\textrm{th}}$
relay of the $q^{\textrm{th}}$ relay cluster. Then the power constraint
$P_{q,j}$ for the $j^{\textrm{th}}$ relay in the $q^{\textrm{th}}$
relay cluster can be expressed in the form of the general multiple
linear constraints in P1 as $\sum_{l=1}^{L}\textrm{Tr}\left(\mathbf{\Sigma}_{l}\mathbf{\hat{W}}_{n}^{l}\right)+\sum_{q^{'}=1}^{Q}\textrm{Tr}\left(\mathbf{\mathbf{\Sigma}}_{q^{'}}^{R}\mathbf{\hat{W}}_{n,q^{'}}\right)\leq\delta_{n}$
by setting $\delta_{n}=P_{q,j}$, $\mathbf{\hat{W}}_{n}^{l}=\mathbf{0},\forall l$,
$\mathbf{\hat{W}}_{n,q^{'}}=\mathbf{0},\forall q^{'}\neq q$, and
$\mathbf{\hat{W}}_{n,q}=\textrm{BlockDiag}\left[\mathbf{0}_{1},...,\mathbf{0}_{j-1},\mathbf{I}_{j},\mathbf{0}_{j+1},...,\mathbf{0}_{n_{q}}\right]$,
where $\mathbf{I}_{j}$ is an $L_{q,j}\times L_{q,j}$ identity matrix
and $\mathbf{0}_{j^{'}},\forall j^{'}\neq j$ is an $L_{q,j^{'}}\times L_{q,j^{'}}$
zero matrix.
\end{rem}

It is well known that the problems in (\ref{eq:P1}) and (\ref{eq:P2})
are usually non-convex problems with non-zero duality gap. Hence,
the standard Lagrange dual method (LDM) \cite{Boyd_04Book_Convex_optimization}
cannot be used to solve these problems. In \cite{Liu_11sTSP_MLC_localLDM},
we proposed a \textit{local LDM} to find a stationary point for a
non-convex problem. We apply the local LDM to obtain the following
algorithm for problem (\ref{eq:GeneralPF}). 

\smallskip{}

\textit{Algorithm LLDM} (for Finding a Stationary Point of Problem
(\ref{eq:GeneralPF})): 

\textbf{\small Initialization}{\small : Choose initial $\mathbf{\lambda}^{(0)}$
such that $\mathbf{\lambda}_{n}^{(0)}>0,\:\forall n$. Let $i=1$.}{\small \par}

\textbf{\small Step 1}{\small{} (Primal update in the inner loop):
For fixed $\mathbf{\lambda}=\mathbf{\lambda}^{(i-1)}$, find a stationary
point $\mathbf{\bar{\Sigma}}^{\left(i-1\right)},\mathbf{\bar{F}}^{(i-1)}$
of
\begin{eqnarray}
\underset{\mathbf{\mathbf{\Sigma}},\mathbf{F}}{\textrm{max}} & L_{\mathbf{\lambda}}\left(\mathbf{\Sigma},\mathbf{F}\right) & \triangleq L\left(\mathbf{\Sigma},\mathbf{F},\mathbf{\lambda}\right),\:\textrm{s.t.}\:\mathbf{\mathbf{\Sigma}}_{l}\succeq0,\:\forall l,\label{eq:mainpro}
\end{eqnarray}
using some primal algorithm with initial point}%
\footnote{{\small Note that $\mathbf{\bar{\Sigma}}^{(-1)},\mathbf{\bar{F}}^{(-1)}$
is randomly generated.}%
}{\small{} $\mathbf{\bar{\Sigma}}^{(i-2)},\mathbf{\bar{F}}^{(i-2)}$.}{\small \par}

\textbf{\small Step 2}{\small{} (Dual update in the outer loop): Update
$\mathbf{\lambda}$ as
\begin{equation}
\mathbf{\lambda}^{(i)}=\mathbf{\lambda}^{(i-1)}+t^{(i)}\mathbf{z}^{(i)},\label{eq:sub-gradient}
\end{equation}
where $t^{(i)}$ is the step size, and $\mathbf{z}^{(i)}$ is the
optimal solution of the following quadratic programming problem: }{\small \par}

{\small 
\begin{equation}
\underset{\mathbf{z}}{\textrm{min}}\:\mathbf{z}^{T}\mathbf{g}^{(i)}+\frac{1}{2}\mathbf{z}^{T}\mathbf{J}^{(i)}\mathbf{z},\:\textrm{s.t.}\:\mathbf{\lambda}^{(i-1)}+\mathbf{z}\geq0,\label{eq:Optdirec}
\end{equation}
where $\mathbf{g}^{(i)}=\left[g_{1}^{(i)},...,g_{N}^{(i)}\right]^{T}$
with $g_{n}^{(i)}=g_{n}\left(\mathbf{\bar{\Sigma}}^{(i-1)},\mathbf{\bar{F}}^{(i-1)}\right),\:\forall n$,
and $\mathbf{J}^{(i)}\in\mathbb{R}^{N\times N}$ is positive definite. }{\small \par}

\textbf{\small Return to Step 1 until convergence.}\smallskip{}

\subsubsection*{{\small Choice of the Matrix $\mathbf{J}^{(i)}$}}

{\small $\mathbf{J}^{(i)}$ is obtained by the well known BFGS update
as follows \cite{Luenberger_Book08_LNLProgramming} 
\[
\mathbf{J}^{(i+1)}=\begin{cases}
\mathbf{J}^{(i)}+\frac{\mathbf{q}_{i}\mathbf{q}_{i}^{T}}{\mathbf{q}_{i}^{T}\mathbf{p}_{i}}-\frac{\mathbf{J}^{(i)}\mathbf{p}_{i}\mathbf{p}_{i}^{T}\mathbf{J}^{(i)}}{\mathbf{p}_{i}^{T}\mathbf{J}^{(i)}\mathbf{p}_{i}}, & \mathbf{q}_{i}^{T}\mathbf{p}_{i}>0,\\
\mathbf{J}^{(i)}, & \textrm{otherwise},
\end{cases}
\]
\[
\mathbf{p}_{i}=\mathbf{\lambda}^{(i)}-\mathbf{\lambda}^{(i-1)},\:\mathbf{q}_{i}=\mathbf{g}^{(i+1)}-\mathbf{g}^{(i)}.
\]
The initial $\mathbf{J}^{(1)}=2^{m_{b}}\mathbf{I}$, where $m_{b}$
is the smallest positive integer such that $\underset{n}{\textrm{max}}\left|\tilde{z}_{n}\right|<0.5$,
and $\tilde{\mathbf{z}}=\left[\tilde{z}_{1},...,\tilde{z}_{N}\right]$
is the optimal solution of the problem $\underset{\mathbf{z}}{\textrm{max}}\:\mathbf{z}^{T}\mathbf{g}^{(1)}+\frac{1}{2}\mathbf{z}^{T}\mathbf{J}^{(1)}\mathbf{z}$.}{\small \par}

\subsubsection*{{\small Choice of the Step Size $t^{(i)}$}}

{\small Set $t^{(i)}=\alpha^{(i)}2^{-m_{t}}$, where $m_{t}$ is an
integer which is initialized as $0$ and is incremented until one
of the following conditions is satisfied: 1) $L\left(\mathbf{\Sigma}^{(i)},\mathbf{F}^{(i)},\mathbf{\lambda}^{(i)}\right)\leq L\left(\mathbf{\Sigma}^{(i-1)},\mathbf{F}^{(i-1)},\mathbf{\lambda}^{(i-1)}\right)$.
2) $m_{t}=m_{t}^{0}$, where $m_{t}^{0}$ is a small positive integer,
e.g., we let $m_{t}^{0}=2$ in the simulations. 3) $re_{i}\leq re_{i-1}$,
where 
\begin{equation}
re_{i}=\underset{n}{\textrm{max}}\left|\lambda_{n}^{(i)}g_{n}\left(\mathbf{\bar{\Sigma}}^{(i)},\mathbf{\bar{F}}^{(i)}\right)\right|+\left(\underset{n}{\textrm{max}}\left\{ g_{n}\left(\mathbf{\bar{\Sigma}}^{(i)},\mathbf{\bar{F}}^{(i)}\right)\right\} \right)^{+},\label{eq:rerr}
\end{equation}
is defined as the }\textit{\small residual error}{\small{} after the
$i^{\textrm{th}}$ iteration. Finally, $\alpha^{(i)}$ is given by
\[
\alpha^{(i+1)}=\left(1-\beta\right)\alpha^{(i)}+\beta t^{(i)},
\]
where $0<\alpha^{(0)}\leq0.5$, $0<\beta<1$ and are set as $\alpha^{(0)}=0.25,\beta=0.2$
in the simulations.}{\small \par}

Algorithm LLDM converges to a stationary point of Problem (\ref{eq:GeneralPF})
under mild conditions \cite{Liu_11sTSP_MLC_localLDM}. In the subsequent
sections, we first characterize the PWF structure at a stationary
point of Problem (\ref{eq:mainpro}). Then we design several efficient
primal algorithms based on the PWF and duality.

\subsection{Structural Properties of a Stationary Point\label{sub:PWF_Structure}}

At any stationary point of Problem (\ref{eq:mainpro}) with fixed
relay precoding matrices $\mathbf{F}$, the input covariance matrices
$\mathbf{\Sigma}$ must satisfy a PWF structure as stated in the following
corollary. 
\begin{cor}
[PWF for fixed relay precoding]\label{cor:PWF_MLC} Let $\mathbf{\bar{\Sigma}}$
be a stationary point of problem (\ref{eq:mainpro}) with fixed $\mathbf{F}$,
i.e., $\mathbf{\bar{\Sigma}}$ satisfies the following KKT conditions
\begin{eqnarray}
\nabla_{\mathbf{\Sigma}_{l}}L_{\lambda}\left(\mathbf{\bar{\Sigma}},\mathbf{F}\right) & \preceq & 0,\:\forall l,\nonumber \\
\textrm{Tr}\left(\mathbf{\bar{\Sigma}}_{l}\nabla_{\mathbf{\Sigma}_{l}}L_{\lambda}\left(\mathbf{\bar{\Sigma}},\mathbf{F}\right)\right) & = & 0,\:\forall l.\label{eq:KKTsigma}
\end{eqnarray}
Then the Type I dual transformation of $\mathbf{\bar{\Sigma}},\mathbf{F}$
is $\bar{\hat{\mathbf{\Sigma}}},\mathbf{\hat{F}}=\left(\mathbf{F}_{1}^{\dagger},...,\mathbf{F}_{Q}^{\dagger}\right)$,
where $\bar{\hat{\mathbf{\Sigma}}}=\left(\mathbf{\bar{\hat{\Sigma}}}_{1},...,\mathbf{\bar{\hat{\Sigma}}}_{L}\right)$
with 
\begin{equation}
\mathbf{\bar{\hat{\Sigma}}}_{l}=w_{l}\left(\mathbf{\bar{\Omega}}_{l}^{-1}-\left(\mathbf{\bar{\Omega}}_{l}+\mathbf{\breve{H}}_{l,l}\bar{\mathbf{\Sigma}}_{l}\mathbf{\check{H}}_{l,l}^{\dagger}\right)^{-1}\right),\:\forall l.\label{eq:sigmhead1}
\end{equation}
Obtain the interference-plus-noise covariance matrices $\mathbf{\bar{\Omega}}_{l}$'s
from (\ref{eq:whiteMG}) using $\mathbf{\bar{\Sigma}},\mathbf{F}$
and the dual ones $\mathbf{\bar{\hat{\Omega}}}_{l}$'s from (\ref{eq:whiteMG-1})
using $\bar{\hat{\mathbf{\Sigma}}},\mathbf{\hat{F}}$. For each $l$,
create an equivalent channel: $\bar{\mathbf{H}}_{l}=\mathbf{\bar{\Omega}}_{l}^{-1/2}\mathbf{\check{H}}_{l,l}\bar{\hat{\mathbf{\Omega}}}_{l}^{-1/2}$.
Perform the thin SVD $\bar{\mathbf{H}}_{l}=\mathbf{E}_{l}\mathbf{\Delta}_{l}\mathbf{G}_{l}^{\dagger}$,
where $\mathbf{E}_{l}\in\mathbb{C}^{L_{R_{l}}\times N_{l}},\ \mathbf{G}_{l}\in\mathbb{C}^{L_{T_{l}}\times N_{l}},\ \mathbf{\Delta}_{l}\in\mathbb{R}_{++}^{N_{l}\times N_{l}}$,
and $N_{l}=\textrm{Rank}\left(\mathbf{\check{H}}_{l,l}\right)$. Then,
$\mathbf{\bar{\Sigma}}$ must have a polite water-filling structure,
i.e.,
\begin{equation}
\mathbf{Q}_{l}\triangleq\bar{\hat{\mathbf{\Omega}}}_{l}^{1/2}\bar{\mathbf{\Sigma}}_{l}\bar{\hat{\mathbf{\Omega}}}_{l}^{1/2}=\mathbf{G}_{l}\left(w_{l}\mathbf{I}-\mathbf{\Delta}_{l}^{-2}\right)^{+}\mathbf{G}_{l}^{\dagger},\:\forall l.\label{eq:PWF_MGL}
\end{equation}
Furthermore, $\bar{\hat{\mathbf{\Sigma}}}$ also possesses the polite
water-filling structure, i.e., 
\begin{equation}
\mathbf{\hat{Q}}_{l}\triangleq\mathbf{\bar{\Omega}}_{l}^{1/2}\bar{\hat{\mathbf{\Sigma}}}_{l}\mathbf{\bar{\Omega}}_{l}^{1/2}=\mathbf{E}_{l}\left(w_{l}\mathbf{I}-\mathbf{\Delta}_{l}^{-2}\right)^{+}\mathbf{E}_{l}^{\dagger},\:\forall l.\label{eq:PWF_MLC_dual}
\end{equation}
On the other hand, if certain $\mathbf{\Sigma}$ has the above polite
water-filling structure for a given $\mathbf{F}$, it must be a stationary
point of problem (\ref{eq:mainpro}) for this fixed $\mathbf{F}$.\hfill \QED
\end{cor}

Corollary \ref{cor:PWF_MLC} follows straightforward from the PWF
results for B-MAC IFN in Theorem 3 of \cite{Liu_10sTSP_MLC} and the
network equivalence result in Theorem \ref{thm:Network-Equivalence}.
The detailed proof is omitted due to limited space.

In some cases, better algorithms can be designed by considering the
following \textit{dual network problem}.
\begin{defn}
[Dual network problem]The \textit{dual network problem} of (\ref{eq:mainpro})
is defined as
\begin{equation}
\underset{\mathbf{\mathbf{\hat{\Sigma}}},\mathbf{\hat{F}}}{\textrm{max}}\:\hat{L}_{\lambda}\left(\mathbf{\hat{\Sigma}},\mathbf{\hat{F}}\right)\triangleq\sum_{l=1}^{L}w_{l}\mathcal{\hat{I}}_{l}^{\Phi^{T}}\left(\mathbf{\mathbf{\hat{\Sigma}}},\mathbf{\hat{F}}\right)-\sum_{l=1}^{L}\textrm{Tr}\left(\mathbf{\hat{\mathbf{\Sigma}}}_{l}\mathbf{W}_{Q+1}^{l}\right)-\sum_{q=1}^{Q}\textrm{Tr}\left(\mathbf{\mathbf{\hat{\Sigma}}}_{q}^{R}\mathbf{W}_{q}\right),\:\textrm{s.t.}\:\mathbf{\mathbf{\hat{\Sigma}}}_{l}\succeq0,\:\forall l.\label{eq:mainpro_dual}
\end{equation}
\end{defn}
\begin{thm}
[Equivalence of problem (\ref{eq:mainpro}) and its dual]\label{thm:dualnetPro}
For any stationary point $\mathbf{\mathbf{\bar{\hat{\Sigma}}}},\mathbf{\bar{\hat{F}}}$
of dual network problem (\ref{eq:mainpro_dual}), its Type I dual
transformation $\mathbf{\bar{\Sigma}},\mathbf{\bar{F}}$ must also
be a stationary point of problem (\ref{eq:mainpro}).
\end{thm}

Please refer to Appendix \ref{sub:Proof-of-Theoremdualpro} for the
proof.

\subsection{Primal Algorithms for two-hop B-MAC AF Relay networks}

We present several primal algorithms for two-hop B-MAC AF Relay networks.
In this case, the relay precoding matrices is $\mathbf{F}=\textrm{BlockDiag}\left[\mathbf{F}\left(1\right),...,\mathbf{F}\left(n_{1}\right)\right]$,
where $\mathbf{F}\left(j\right)\in\mathbb{C}^{L_{1,j}\times L_{1,j}}$
is the relay precoding matrix at the $j^{\textrm{th}}$ relay, $L_{1,j}$
is the number of antennas at the $j^{\textrm{th}}$ relay, and $n_{1}$
is the number of relays.

\subsubsection{Gradient Ascent Primal Algorithm\label{sub:Baseline-Algorithm-(Gradient} }

The gradient ascent (GA) algorithm with the step size determined by
Armijo rule \cite{Bertsekas_book99_NProgramming} can be used to find
a stationary point of Problem (\ref{eq:mainpro}) as summarized in
Table \ref{tab:AlgGP}. In Algorithm GA, the gradient update for source
node is performed over the precoding matrices $\mathbf{T}_{l}$'s,
where $\mathbf{T}_{l}\mathbf{T}_{l}^{\dagger}=\mathbf{\Sigma}_{l},\:\forall l$.
For convenience, we group the gradients over $\mathbf{F}\left(j\right),j=1,...,n_{1}$
as a block diagonal matrix $\nabla_{\mathbf{F}(1:n_{1})}\left(\mathbf{\Sigma},\mathbf{F}\right)\triangleq\textrm{BlockDiag}\left[\nabla_{\mathbf{F}(1)}L_{\lambda}\left(\mathbf{\Sigma},\mathbf{F}\right),...,\nabla_{\mathbf{F}(n_{1})}L_{\lambda}\left(\mathbf{\Sigma},\mathbf{F}\right)\right]$.
Then the gradients required in Algorithm GA are given by \cite{Hjorungnes_TSP07_ComplexDiff}
\begin{equation}
\nabla_{\mathbf{T}_{l}}L_{\lambda}\left(\mathbf{\Sigma},\mathbf{F}\right)=-2\hat{\mathbf{W}}_{l}^{'}\mathbf{T}_{l}+2w_{l}\mathbf{\check{H}}_{l,l}^{\dagger}\mathbf{X}_{l}^{-1}\mathbf{\check{H}}_{l,l}\mathbf{T}_{l}-2\sum_{k\neq l}w_{k}\phi_{k,l}\mathbf{\check{H}}_{k,l}^{\dagger}\left(\mathbf{\Omega}_{k}^{-1}-\mathbf{X}_{k}^{-1}\right)\mathbf{\check{H}}_{k,l}\mathbf{T}_{l},\:\forall l,\label{eq:Gradsigma}
\end{equation}
\begin{equation}
\nabla_{\mathbf{F}(1:n_{1})}\left(\mathbf{\Sigma},\mathbf{F}\right)=\textrm{BlockDiag}\left[\mathbf{1}_{1},...,\mathbf{1}_{n_{1}}\right]\circ\nabla_{\mathbf{F}}L_{\lambda}\left(\mathbf{\Sigma},\mathbf{F}\right),\label{eq:GradF}
\end{equation}
where $\mathbf{X}_{l}=\mathbf{\Omega}_{l}+\mathbf{\check{H}}_{l,l}\mathbf{\Sigma}_{l}\mathbf{\check{H}}_{l,l}^{\dagger},\forall l$;
the notation $\circ$ denotes the pointwise product of two matrices;
$\mathbf{1}_{j},\forall j\in\left\{ 1,...,n_{1}\right\} $ is an $L_{1,j}\times L_{1,j}$
matrix of which all the elements are ones; and $\nabla_{\mathbf{F}}L_{\lambda}\left(\mathbf{\Sigma},\mathbf{F}\right)$
is given by 
\begin{equation}
\nabla_{\mathbf{F}}L_{\lambda}\left(\mathbf{\Sigma},\mathbf{F}\right)=2\sum_{l=1}^{L}w_{l}\left(\mathbf{H}_{1}^{l\dagger}\mathbf{X}_{l}^{-1}\mathbf{A}_{l}-\mathbf{H}_{1}^{l\dagger}\mathbf{\Omega}_{l}^{-1}\mathbf{C}_{l}\right)-2\sum_{l=1}^{L}\mathbf{\hat{W}}_{1}\mathbf{F}\mathbf{H}_{0}^{l}\mathbf{\Sigma}_{l}\mathbf{H}_{0}^{l\dagger}-2\mathbf{\hat{W}}_{1}\mathbf{F}\mathbf{W}_{1},\label{eq:GF}
\end{equation}
where $\mathbf{A}_{l}=\mathbf{H}_{1}^{l}\mathbf{F}\left(\sum_{k=1}^{L}\phi_{l,k}\mathbf{H}_{0}^{k}\mathbf{\Sigma}_{k}\mathbf{H}_{0}^{k\dagger}+\right.$
$\left.\mathbf{H}_{0}^{l}\mathbf{\Sigma}_{l}\mathbf{H}_{0}^{l\dagger}+\mathbf{W}_{1}\right)$,
and $\mathbf{C}_{l}=\mathbf{H}_{1}^{l}\mathbf{F}\left(\sum_{k=1}^{L}\phi_{l,k}\mathbf{H}_{0}^{k}\mathbf{\Sigma}_{k}\mathbf{H}_{0}^{k\dagger}+\mathbf{W}_{1}\right)$.%

\begin{table}
\caption{\label{tab:AlgGP}Algorithm GA (Gradient Ascent Method for Finding
a Stationary Point of Problem (\ref{eq:mainpro}))}

\centering{}%
\begin{tabular}{l}
\hline 
{\small Initialize $\mathbf{T}_{l},\:\forall l$ and $\mathbf{F}$=$\textrm{BlockDiag}\left[\mathbf{F}\left(1\right),...,\mathbf{F}\left(n_{1}\right)\right]$.}\tabularnewline
{\small $\;$}\textbf{\small Do}\tabularnewline
{\small $\;$$\;$1. Calculate the gradient over $\mathbf{T}_{l}$'s
using (\ref{eq:Gradsigma}): $\mathbf{G}_{l}^{\textrm{T}}=\nabla_{\mathbf{T}_{l}}L_{\lambda}\left(\mathbf{\Sigma},\mathbf{F}\right),\: l=1,...,L$.}\tabularnewline
{\small $\;$$\;$2. Choose step size $\alpha$ via Armijo rule and
update $\mathbf{T}_{l}$'s as $\mathbf{T}_{l}=\mathbf{T}_{l}+\alpha\mathbf{G}_{l}^{\textrm{T}},\: l=1,...,L$.}\tabularnewline
{\small $\;$$\;$3. Calculate the gradient over $\mathbf{F}\left(j\right)$'s
using (\ref{eq:GradF}): $\mathbf{G}_{j}^{\textrm{F}}=\nabla_{\mathbf{F}(j)}L_{\lambda}\left(\mathbf{\Sigma},\mathbf{F}\right),j=1,...,n_{1}$.}\tabularnewline
{\small $\;$$\;$4. Choose step size $\alpha$ via Armijo rule and
update $\mathbf{F}$ as $\mathbf{F}\left(j\right)=\mathbf{F}\left(j\right)+\alpha\mathbf{G}_{j}^{\textrm{F}},j=1,...,n_{1}$.}\tabularnewline
{\small $\;$}\textbf{\small until}{\small{} all the norms of $\mathbf{G}_{l}^{\textrm{T}},\:\forall l$
and $\mathbf{G}_{j}^{\textrm{F}},\:\forall j$ are small enough}\tabularnewline
\hline 
\end{tabular}
\end{table}

\subsubsection{Polite Water-filling based Primal Algorithm\label{sub:Polite-water-filling-based}}

We exploit the duality and PWF structure to design \textit{Algorithm
PWF} as summarized in Table \ref{tab:AlgPWF}. By Corollary \ref{cor:PWF_MLC}
and the property of gradient update for $\mathbf{F}(j)$'s, if Algorithm
PWF converges, the solution must be a stationary point of problem
(\ref{eq:mainpro}). Simulations show that Algorithm PWF has a faster
convergence speed than Algorithm GA.
\begin{table}
\caption{\label{tab:AlgPWF}Algorithm PWF (PWF based Algorithm for Finding
a Stationary Point of Problem (\ref{eq:mainpro}))}

\centering{}%
\begin{tabular}{l}
\hline 
{\small Initialize $\mathbf{\Sigma}$ and $\mathbf{F}=\textrm{BlockDiag}\left[\mathbf{F}\left(1\right),...,\mathbf{F}\left(n_{1}\right)\right]$
such that $\mathbf{\Sigma}_{l}\succeq0,\forall l$. }\tabularnewline
{\small Obtain the initial $\mathbf{\hat{\Sigma}}$ from (\ref{eq:sigmhead1})
using the initial $\mathbf{\Sigma}$. }\tabularnewline
\textbf{\small While}{\small{} not converge }\textbf{\small do}{\small{} }\tabularnewline
{\small $\;$1. For fixed $\mathbf{\hat{\Sigma}}$ and $\mathbf{F}$,
update $\mathbf{\Sigma}$ by polite water-filling:}\tabularnewline
{\small $\;$$\;$Obtain $\mathbf{\hat{\Omega}}_{l}=\mathbf{\hat{W}}_{l}^{'}+\sum_{k=1}^{L}\mathbf{\Phi}_{k,l}\mathbf{\check{H}}_{k,l}^{\dagger}\mathbf{\hat{\Sigma}}_{k}\mathbf{\check{H}}_{k,l},\:\forall l$
where $\hat{\mathbf{W}}_{l}^{'}$ is given in (\ref{eq:Whlp}).}\tabularnewline
{\small $\;$$\;$}\textbf{\small For}{\small{} $l=1$ to $L$}\tabularnewline
{\small $\;$$\;$$\;$a. Obtain $\mathbf{\Omega}_{l}$ using (\ref{eq:whiteMG})}\tabularnewline
{\small $\;$$\;$$\;$b. Perform thin SVD $\mathbf{\Omega}_{l}^{-1/2}\mathbf{\check{H}}_{l,l}\hat{\mathbf{\Omega}}_{l}^{-1/2}=\mathbf{E}_{l}\mathbf{\Delta}_{l}\mathbf{G}_{l}^{\dagger},\:\forall l$.}\tabularnewline
{\small $\;$$\;$$\;$c. Update $\mathbf{\Sigma}_{l}$ as $\mathbf{\Sigma}_{l}=\mathbf{\hat{\Omega}}_{l}^{-1/2}\mathbf{G}_{l}\left(w_{l}\mathbf{I}-\mathbf{\Delta}_{l}^{-2}\right)^{+}\mathbf{G}_{l}^{\dagger}\mathbf{\hat{\Omega}}_{l}^{-1/2}$.}\tabularnewline
{\small $\;$$\;$}\textbf{\small End}\tabularnewline
{\small $\;$2. For fixed $\mathbf{\Sigma}$, update $\mathbf{F}$
as $\mathbf{F}\left(j\right)=\mathbf{F}\left(j\right)+\alpha\nabla_{\mathbf{F}(j)}L_{\lambda}\left(\mathbf{\Sigma},\mathbf{F}\right),j=1,...,n_{1}$,}\tabularnewline
{\small $\;$$\;$$\;$$\;$$\;$where the step size $\alpha$ is
obtained by Armijo rule.}\tabularnewline
{\small $\;$3. For fixed $\mathbf{\Sigma}$ and $\mathbf{F}$, update
$\mathbf{\hat{\Sigma}}$ by polite water-filling:}\tabularnewline
{\small $\;$$\;$Obtain $\mathbf{\Omega}_{l},\:\forall l$ using
(\ref{eq:whiteMG}).}\tabularnewline
{\small $\;$$\;$}\textbf{\small For}{\small{} $l=1$ to $L$}\tabularnewline
{\small $\;$$\;$$\;$a. obtain $\mathbf{\hat{\Omega}}_{l}=\mathbf{\hat{W}}_{l}^{'}+\sum_{k=1}^{L}\mathbf{\Phi}_{k,l}\mathbf{\check{H}}_{k,l}^{\dagger}\mathbf{\hat{\Sigma}}_{k}\mathbf{\check{H}}_{k,l},$
where $\hat{\mathbf{W}}_{l}^{'}$ is given in (\ref{eq:Whlp}).}\tabularnewline
{\small $\;$$\;$$\;$b. Perform thin SVD $\mathbf{\Omega}_{l}^{-1/2}\mathbf{\check{H}}_{l,l}\hat{\mathbf{\Omega}}_{l}^{-1/2}=\mathbf{E}_{l}\mathbf{\Delta}_{l}\mathbf{G}_{l}^{\dagger},\:\forall l$.}\tabularnewline
{\small $\;$$\;$$\;$c. Update $\mathbf{\hat{\Sigma}}_{l}$ as $\mathbf{\hat{\Sigma}}_{l}=\mathbf{\Omega}_{l}^{-1/2}\mathbf{E}_{l}\left(w_{l}\mathbf{I}-\mathbf{\Delta}_{l}^{-2}\right)^{+}\mathbf{E}_{l}^{\dagger}\mathbf{\Omega}_{l}^{-1/2}$.}\tabularnewline
{\small $\;$$\;$}\textbf{\small End}\tabularnewline
\textbf{\small End}\tabularnewline
\hline 
\end{tabular}
\end{table}

\subsubsection{Structured Primal Algorithm for two-hop BC AF relay networks\label{sub:AlgMACBC}}

Consider sum rate maximization under multiple linear constraints (i.e.,
problem (\ref{eq:P1}) with $\mu_{l}=1,\:\forall l$) in a two-hop
BC AF relay network applying DPC. We apply the duality and Corollary
\ref{cor:PWF_MLC} to design a structured primal algorithm to find
a stationary point of the inner loop problem (\ref{eq:mainpro}) for
this special case.

Note that in this case, $Q=1$, $\mathbf{H}_{\textrm{0}}^{l},\:\forall l$
are the same, and $\mathbf{\hat{W}}_{0}^{l},\:\forall l$ are the
same. For convenience, let
\begin{equation}
\left(\mathbf{H}_{\textrm{b}},\left[\mathbf{H}_{\textrm{b}}^{l}\right],\mathbf{W}_{\textrm{b}}^{\textrm{r}},\left[\mathbf{W}_{\textrm{b}}^{l}\right],\mathbf{\hat{W}}_{\textrm{b}},\mathbf{\hat{W}}_{\textrm{b}}^{\textrm{r}}\right),\label{eq:BCrelay}
\end{equation}
denote a two-hop BC AF relay network where the channel matrix between
source and relay is $\mathbf{H}_{0}^{l}=\mathbf{H}_{\textrm{b}},\:\forall l$,
the channel matrix between relay and the $l^{\textrm{th}}$ destination
is $\mathbf{H}_{Q}^{l}=\mathbf{H}_{\textrm{b}}^{l}$, the covariance
of the noise at the relay is $\mathbf{W}_{1}=\mathbf{W}_{\textrm{b}}^{\textrm{r}}$,
the covariance of the noise at the $l^{\textrm{th}}$ destination
is $\mathbf{W}_{2}^{l}=\mathbf{W}_{\textrm{b}}^{l}$, and the constraint
matrices in Lagrangian (\ref{eq:Lag}) is $\mathbf{\hat{W}}_{0}^{l}=\mathbf{\hat{W}}_{\textrm{b}},\:\forall l$,
$\hat{\mathbf{W}}_{1}=\mathbf{\hat{W}}_{\textrm{b}}^{\textrm{r}}$.
The dual network of (\ref{eq:BCrelay}) is
\begin{equation}
\left(\left[\mathbf{H}_{\textrm{m}}^{l}\right],\mathbf{H}_{\textrm{m}},\mathbf{W}_{\textrm{m}}^{\textrm{r}},\mathbf{W}_{\textrm{m}},\left[\mathbf{\hat{W}}_{\textrm{m}}^{l}\right],\mathbf{\hat{W}}_{\textrm{m}}^{\textrm{r}}\right),\label{eq:MACrelay}
\end{equation}
which is a two-hop MAC AF relay network where the channel matrix between
the $l^{\textrm{th}}$ source and relay is $\mathbf{H}_{\textrm{m}}^{l}=\mathbf{H}_{\textrm{b}}^{l\dagger},\:\forall l$,
the channel matrix between relay and the destination is $\mathbf{H}_{\textrm{m}}=\mathbf{H}_{\textrm{b}}^{\dagger}$,
the covariance of the noise at the relay is $\mathbf{W}_{\textrm{m}}^{\textrm{r}}=\mathbf{\hat{W}}_{\textrm{b}}^{\textrm{r}}$,
the covariance of the noise at the $l^{\textrm{th}}$ destination
is $\mathbf{W}_{\textrm{m}}=\mathbf{\hat{W}}_{\textrm{b}}$, and the
constraint matrices in Lagrangian (\ref{eq:Lag}) is $\mathbf{\hat{W}}_{\textrm{m}}^{l}=\mathbf{W}_{\textrm{b}}^{l},\:\forall l$,
$\mathbf{\hat{W}}_{\textrm{m}}^{\textrm{r}}=\mathbf{W}_{\textrm{b}}^{\textrm{r}}$.
By Theorem \ref{thm:dualnetPro}, Problem (\ref{eq:mainpro}) for
the BC relay network (\ref{eq:BCrelay}) can be equivalently solved
by solving the dual network problem (\ref{eq:mainpro_dual}), whose
objective function, in this special case, is given by 
\begin{eqnarray}
\hat{L}_{\lambda}\left(\mathbf{\hat{\Sigma}},\mathbf{\hat{F}}\right) & = & \textrm{log}\left|\frac{\mathbf{H}_{\textrm{m}}\mathbf{\hat{F}}\mathbf{R}\mathbf{\mathbf{\hat{F}}}^{\dagger}\mathbf{H}_{\textrm{m}}^{\dagger}+\mathbf{H}_{\textrm{m}}\mathbf{\hat{F}}\mathbf{W}_{\textrm{m}}^{\textrm{r}}\mathbf{\mathbf{\hat{F}}}^{\dagger}\mathbf{H}_{\textrm{m}}^{\dagger}+\mathbf{W}_{\textrm{m}}}{\mathbf{H}_{\textrm{m}}\mathbf{\hat{F}}\mathbf{W}_{\textrm{m}}^{\textrm{r}}\mathbf{\mathbf{\hat{F}}}^{\dagger}\mathbf{H}_{\textrm{m}}^{\dagger}+\mathbf{W}_{\textrm{m}}}\right|\label{eq:dualLag}\\
 &  & -\textrm{Tr}\left(\mathbf{\hat{F}}\left(\mathbf{R}+\mathbf{W}_{\textrm{m}}^{\textrm{r}}\right)\mathbf{\hat{F}}^{\dagger}\mathbf{\hat{W}}_{\textrm{m}}^{\textrm{r}}\right)-\sum_{l=1}^{L}\textrm{Tr}\left(\mathbf{\mathbf{\hat{\Sigma}}}_{l}\mathbf{\hat{W}}_{\textrm{m}}^{l}\right),\nonumber 
\end{eqnarray}
where $\mathbf{R}=\sum_{l=1}^{L}\mathbf{H}_{\textrm{m}}^{l}\mathbf{\hat{\Sigma}}_{l}\mathbf{H}_{\textrm{m}}^{l\dagger}$.

By exploiting the specific structure of $\hat{L}_{\lambda}$, we propose
an improved algorithm over PWF.\smallskip{}

\textit{Algorithm PWFI}: 

\textbf{\small Initialization}{\small : Choose proper initial $\hat{\mathbf{\Sigma}},\mathbf{\hat{F}}$
such that $\mathbf{\hat{\Sigma}}_{l}\succeq0,\:\forall l$. }{\small \par}

\textbf{\small Step }{\small 1: For fixed $\mathbf{\hat{F}}$ and
$\mathbf{\hat{\Sigma}}_{k},\:\forall k\neq l$, Problem (\ref{eq:mainpro_dual})
reduces to the single-user optimization problem:
\begin{equation}
\underset{\mathbf{\hat{\Sigma}}}{\textrm{max}}\:\textrm{log}\left|\mathbf{\bar{H}}_{l}\mathbf{\hat{\Sigma}}_{l}\mathbf{\bar{H}}_{l}^{\dagger}+\mathbf{\bar{W}}_{l}\right|-\textrm{Tr}\left(\mathbf{\mathbf{\hat{\Sigma}}}_{l}\mathbf{\bar{\hat{W}}}_{l}\right),\label{eq:maxLagl}
\end{equation}
where $\mathbf{\bar{H}}_{l}=\mathbf{H}_{\textrm{m}}\mathbf{\hat{F}}\mathbf{H}_{\textrm{m}}^{l}$,
$\mathbf{\bar{W}}_{l}=\mathbf{H}_{\textrm{m}}\mathbf{\hat{F}}\sum_{k\neq l}\mathbf{H}_{\textrm{m}}^{k}\mathbf{\hat{\Sigma}}_{k}\mathbf{H}_{\textrm{m}}^{k\dagger}\mathbf{\mathbf{\hat{F}}}^{\dagger}\mathbf{H}_{\textrm{m}}^{\dagger}+\mathbf{H}_{\textrm{m}}\mathbf{\hat{F}}\mathbf{W}_{\textrm{m}}^{\textrm{r}}\mathbf{\mathbf{\hat{F}}}^{\dagger}\mathbf{H}_{\textrm{m}}^{\dagger}+\mathbf{W}_{\textrm{m}}$,
and $\mathbf{\bar{\hat{W}}}_{l}=\mathbf{H}_{\textrm{m}}^{l\dagger}\mathbf{\hat{F}}^{\dagger}\mathbf{\hat{W}}_{\textrm{m}}^{\textrm{r}}\hat{\mathbf{F}}\mathbf{H}_{\textrm{m}}^{l}+\mathbf{\hat{W}}_{\textrm{m}}^{l}$.
Applying Corollary \ref{cor:PWF_MLC} to the single-user optimization
problem in (\ref{eq:maxLagl}), the optimal solution is given by 
\begin{equation}
\mathbf{\mathbf{\hat{\Sigma}}}_{l}=\mathbf{\bar{\hat{W}}}_{l}^{-1/2}\mathbf{G}_{l}\left(\mathbf{I}-\mathbf{\Delta}_{l}^{-2}\right)^{+}\mathbf{G}_{l}^{\dagger}\mathbf{\bar{\hat{W}}}_{l}^{-1/2},\:\forall l,\label{eq:GWFsolution}
\end{equation}
where $\mathbf{G}_{l}$ and $\mathbf{\Delta}_{l}$ are obtained by
performing the thin SVD $\mathbf{\bar{W}}_{l}^{-1/2}\mathbf{\bar{H}}_{l}\mathbf{\bar{\hat{W}}}_{l}^{-1/2}=\mathbf{E}_{l}\mathbf{\Delta}_{l}\mathbf{G}_{l}^{\dagger}$.
Update each $\mathbf{\mathbf{\hat{\Sigma}}}_{l}$ for once using (\ref{eq:GWFsolution})
when fixing $\mathbf{\hat{F}}$ and $\mathbf{\hat{\Sigma}}_{k},\:\forall k\neq l$.}{\small \par}

\textbf{\small Step 2}{\small : For fixed $\mathbf{\hat{\Sigma}}$,
$\mathbf{\hat{F}}$ is updated by solving $\underset{\mathbf{\hat{F}}}{\textrm{max}}\:\hat{L}_{\lambda}\left(\mathbf{\hat{\Sigma}},\mathbf{\hat{F}}\right)$,
which is equivalent to the following problem
\begin{equation}
\underset{\mathbf{\bar{F}}}{\textrm{max}}\:\textrm{log}\left|\frac{\mathbf{\bar{H}}^{\dagger}\mathbf{\bar{F}}^{\dagger}\mathbf{\bar{R}}\mathbf{\bar{F}}\mathbf{\bar{H}}+\mathbf{\bar{H}}^{\dagger}\mathbf{\bar{F}}^{\dagger}\mathbf{\bar{F}}\mathbf{\bar{H}}+\mathbf{I}}{\mathbf{\bar{H}}^{\dagger}\mathbf{\bar{F}}^{\dagger}\mathbf{\bar{F}}\mathbf{\bar{H}}+\mathbf{I}}\right|-\textrm{Tr}\left(\mathbf{\bar{F}}^{\dagger}\left(\mathbf{\bar{R}}+\mathbf{I}\right)\mathbf{\bar{F}}\right),\label{eq:MACP}
\end{equation}
where $\mathbf{\bar{H}}=\left(\mathbf{\hat{W}}_{\textrm{m}}^{\textrm{r}}\right)^{-1/2}\mathbf{H}_{\textrm{m}}^{\dagger}\mathbf{W}_{\textrm{m}}^{-1/2}$,
$\mathbf{\bar{F}}=\left(\mathbf{W}_{\textrm{m}}^{\textrm{r}}\right)^{1/2}\mathbf{\hat{F}}^{\dagger}\left(\mathbf{\hat{W}}_{\textrm{m}}^{\textrm{r}}\right)^{1/2}$
and $\mathbf{\bar{R}}=\left(\mathbf{W}_{\textrm{m}}^{\textrm{r}}\right)^{-1/2}\mathbf{R}\left(\mathbf{W}_{\textrm{m}}^{\textrm{r}}\right)^{-1/2}$.
Define $M=\textrm{min}\left(\textrm{Rank}\left(\mathbf{\bar{H}}\right),\textrm{Rank}\left(\mathbf{\bar{R}}\right)\right)$.
Perform the SVD $\mathbf{\bar{H}}=\mathbf{U}_{h}\mathbf{D}_{h}\mathbf{V}_{h}$
and the eigenvalue decomposition $\mathbf{\bar{R}}=\mathbf{U}_{r}\mathbf{D}_{r}\mathbf{V}_{r}$.
Let $\sigma_{1},...,\sigma_{M}$ denote the $M$ largest singular
values of $\mathbf{\bar{H}}$ with descending order and $\mathbf{U}_{h}^{M}$
denote the semi-unitary matrix formed by the singular vectors corresponding
to $\sigma_{i}$'s. Let $\delta_{1},...,\delta_{M}$ denote the $M$
largest eigenvalues of $\mathbf{\bar{R}}$ with descending order and
$\mathbf{U}_{r}^{M}$ denote the semi-unitary matrix formed by the
eigenvectors corresponding to $\delta_{i}$'s. Then the optimal $\hat{\mathbf{F}}$
is given by \cite{Yu_TSP10_GWFRL} 
\begin{equation}
\hat{\mathbf{F}}=\left(\mathbf{\hat{W}}_{\textrm{m}}^{\textrm{r}}\right)^{-1/2}\mathbf{U}_{h}^{M}\mathbf{D}_{f}\mathbf{U}_{r}^{M\dagger}\left(\mathbf{W}_{\textrm{m}}^{\textrm{r}}\right)^{-1/2},\label{eq:updateFhed}
\end{equation}
where $\mathbf{D}_{f}=\textrm{diag}\left(f_{1},...,f_{M}\right)^{1/2}\geq0$
with $f_{i},i=1,...,M$ given by 
\[
f_{i}=\frac{1}{2\sigma_{i}^{2}\left(\delta_{i}+1\right)}\left[\sqrt{\delta_{i}^{2}+4\delta_{i}\sigma_{i}^{2}}-\delta_{i}-2\right]^{+}.
\]
}{\small \par}

\textbf{\small Return to step 1 until convergence.}\smallskip{}

Since the objective $\hat{L}_{\lambda}\left(\mathbf{\hat{\Sigma}},\mathbf{\hat{F}}\right)$
is upper bounded and is increased after each update, Algorithm PWFI
must converge to a fixed point $\mathbf{\bar{\hat{\Sigma}}},\mathbf{\bar{\hat{F}}}$.
It can be shown that the fixed point $\mathbf{\bar{\hat{\Sigma}}},\mathbf{\bar{\hat{F}}}$
must be a stationary point of the dual network problem (\ref{eq:mainpro_dual}).
Then by Theorem \ref{thm:dualnetPro}, the corresponding Type I dual
transformation $\mathbf{\bar{\Sigma}},\mathbf{\bar{F}}$ is a stationary
point of Problem (\ref{eq:mainpro}) in the BC relay network (\ref{eq:BCrelay}).

\subsection{Primal Algorithms for Multi-hop B-MAC AF Relay Networks\label{sub:Algorithm-for-multi-hop}}

Algorithm GA/PWF can be easily extended to solve Problem (\ref{eq:mainpro})
for B-MAC AF relay networks with more than two hops. The only difference
is that the gradient update of the relay precoding matrices is generalized
to multi-hop case as 
\[
\mathbf{F}_{q}\left(j\right)=\mathbf{F}_{q}\left(j\right)+\alpha\nabla_{\mathbf{F}_{q}\left(j\right)}L_{\lambda}\left(\mathbf{\Sigma},\mathbf{F}\right),\: q=1,...,Q,j=1,...,n_{q},
\]
where the gradient $\nabla_{\mathbf{F}_{q}\left(j\right)}L_{\lambda}\left(\mathbf{\Sigma},\mathbf{F}\right),\:\forall q,j$
can be calculated similar to the one in (\ref{eq:GradF}).%

The duality can be used to simplify the the sum rate maximization
under multiple linear constraints for a three-hop BC relay network.
We design \textit{Algorithm PWF3} to find a stationary point of the
inner-loop problem (\ref{eq:mainpro}) for this special case, where
$\mathbf{H}_{0}^{l}=\mathbf{H}_{0},\:\forall l$, $w_{l}=1,\:\forall l$
and $\mathbf{\hat{W}}_{0}^{l}=\mathbf{\hat{W}}_{0},\:\forall l$.
Algorithm PWF3 switches the optimization between the original network
and the dual network so that at each time, we only need to consider
a two-hop network optimization problem solved in Section \ref{sub:AlgMACBC}.\smallskip{}

\textit{Algorithm PWF}3: 

\textbf{\small Initialization}{\small : Choose proper initial $\mathbf{\Sigma},\mathbf{F}=\left(\mathbf{F}_{1},\mathbf{F}_{2}\right)$
such that $\mathbf{\Sigma}_{l}\succeq0,\:\forall l$. }{\small \par}

\textbf{\small Step 1}{\small : For fixed $\mathbf{F}_{1}$, the inner
loop problem reduces to Problem (\ref{eq:mainpro}) in the two-hop
BC AF relay network (\ref{eq:BCrelay}) with $\mathbf{H}_{\textrm{b}}=\mathbf{H}_{1}\mathbf{F}_{1}\mathbf{H}_{0}$,
$\mathbf{H}_{\textrm{b}}^{l}=\mathbf{H}_{Q}^{l},\:\forall l$, $\mathbf{W}_{\textrm{b}}^{\textrm{r}}=\mathbf{H}_{1}\mathbf{F}_{1}\mathbf{W}_{1}\mathbf{F}_{1}^{\dagger}\mathbf{H}_{1}^{\dagger}+\mathbf{W}_{2}$,
$\mathbf{W}_{\textrm{b}}^{l}=\mathbf{W}_{Q+1}^{l},\:\forall l$, $\mathbf{\hat{W}}_{\textrm{b}}=\mathbf{H}_{0}^{\dagger}\mathbf{F}_{1}^{\dagger}\mathbf{\hat{W}}_{1}\mathbf{F}_{1}\mathbf{H}_{0}+\mathbf{\hat{W}}_{0}$
and $\mathbf{\hat{W}}_{\textrm{b}}^{\textrm{r}}=\mathbf{\hat{W}}_{2}$.
Obtain the input covariance and relay precoding matrices $\mathbf{\hat{\Sigma}}_{b}=\left(\mathbf{\hat{\Sigma}}_{b}^{1},...,\mathbf{\hat{\Sigma}}_{b}^{L}\right),\mathbf{\hat{F}}_{b}$
for the dual two-hop MAC AF relay network (\ref{eq:MACrelay}) by
the Type I dual transformation of $\mathbf{\Sigma},\mathbf{F}_{2}$
applied to network (\ref{eq:BCrelay}). Update each $\mathbf{\mathbf{\hat{\Sigma}}}_{b}^{l}$
for once using (\ref{eq:GWFsolution}) when fixing $\mathbf{\hat{F}}_{b}$
and $\mathbf{\hat{\Sigma}}_{b}^{k},\:\forall k\neq l$. Then update
$\mathbf{\hat{F}}_{b}$ using (\ref{eq:updateFhed}) for fixed $\mathbf{\hat{\Sigma}}_{b}$.
Finally, obtain the updated $\mathbf{\Sigma},\mathbf{F}_{2}$ by the
Type I dual transformation of $\mathbf{\hat{\Sigma}}_{b},\mathbf{\hat{F}}_{b}$
applied to network (\ref{eq:MACrelay}).}{\small \par}

\textbf{\small Step 2}{\small : Obtain $\mathbf{\hat{\Sigma}},\mathbf{\hat{F}}$
by the Type I dual transformation of $\mathbf{\Sigma},\mathbf{F}$. }{\small \par}

\textbf{\small Step 3}{\small : For fixed $\mathbf{\hat{F}}_{2}$,
update $\mathbf{\hat{\Sigma}},\mathbf{\hat{F}}_{1}$ by improving
the objective value of the dual network problem (\ref{eq:mainpro_dual}),
which reduces to Problem (\ref{eq:mainpro}) in the two-hop MAC AF
relay network (\ref{eq:MACrelay}) with $\mathbf{H}_{\textrm{m}}^{l}=\mathbf{H}_{1}^{\dagger}\mathbf{\hat{F}}_{2}\mathbf{H}_{Q}^{l\dagger},\:\forall l$,
$\mathbf{H}_{\textrm{m}}=\mathbf{H}_{0}^{\dagger}$, $\mathbf{W}_{\textrm{m}}^{\textrm{r}}=\mathbf{H}_{1}^{\dagger}\mathbf{\hat{F}}_{2}\mathbf{\hat{W}}_{2}\mathbf{\hat{F}}_{2}^{\dagger}\mathbf{H}_{1}+\mathbf{\hat{W}}_{1}$,
$\mathbf{W}_{\textrm{m}}=\mathbf{\hat{W}}_{0}$, $\mathbf{\hat{W}}_{\textrm{m}}^{l}=\mathbf{H}_{Q}^{l}\mathbf{\hat{F}}_{2}^{\dagger}\mathbf{W}_{2}\mathbf{\hat{F}}_{2}\mathbf{H}_{Q}^{l\dagger}+\mathbf{W}_{Q+1}^{l}$,
and $\mathbf{\hat{W}}_{\textrm{m}}^{\textrm{r}}=\mathbf{W}_{1}$.
Treating $\mathbf{\hat{\Sigma}},\mathbf{\hat{F}}_{1}$ as the input
covariance and relay precoding matrices in network (\ref{eq:MACrelay}),
update $\mathbf{\hat{\Sigma}},\mathbf{\hat{F}}_{1}$ using a single
iteration of Algorithm PWFI applied to network (\ref{eq:MACrelay}).}{\small \par}

\textbf{\small Step 4}{\small : Obtain $\mathbf{\Sigma},\mathbf{F}$
by the Type I dual transformation of $\mathbf{\hat{\Sigma}},\mathbf{\hat{F}}$.}{\small \par}

\textbf{\small Return to step 1 until convergence.}\smallskip{}

Using Theorem \ref{thm:typeIduality}, it can be shown that Algorithm
PWF3 monotonically increases the objective after each iteration and
converges to a stationary point of Problem (\ref{eq:mainpro}).
\begin{rem}
\label{rem:CompPWFIFC}For the special case of B-MAC IFN, the duality
and PWF have been established in \cite{Liu_IT10s_Duality_BMAC,Liu_10sTSP_MLC}.
However, we cannot extend the existing results and algorithms trivially
from \cite{Liu_IT10s_Duality_BMAC,Liu_10sTSP_MLC}. We proposed new
techniques, namely the network equivalence technique in Theorem \ref{thm:Network-Equivalence},
and the network dual scaling technique in Theorem \ref{thm:Network-Dual-Scaling},
to generalize the duality and PWF to B-MAC AF relay networks. The
proof of Theorem \ref{thm:Network-Dual-Scaling} is non-trivial. The
algorithm design also relies on the new network equivalence result.
Moreover, Theorem \ref{thm:dualnetPro} (equivalence of problem (\ref{eq:mainpro})
and its dual) is non-trivial and cannot be re-derived from our previous
results in \cite{Liu_IT10s_Duality_BMAC,Liu_10sTSP_MLC}. Based on
Theorem \ref{thm:dualnetPro}, we proposed structured primal algorithms
(PWFI and PWF3) which are entirely different from the single-hop PWF
algorithms in \cite{Liu_IT10s_Duality_BMAC,Liu_10sTSP_MLC}.
\end{rem}

\section{Simulation Results\label{sec:Simulation-Results}}

Block fading channel is assumed and each channel matrix has zero-mean
i.i.d. Gaussian entries with unit variance. The simulation results
in all figures are averaged over 100 random channel realizations.

\subsection{Verify the Convergence of the Proposed Primal Algorithms}

The parameters of Problem (\ref{eq:mainpro}) are set as $w_{l}=1,\:\mathbf{\hat{W}}_{0}^{l}=0.01\mathbf{I},\:\mathbf{\hat{W}}_{q}=0.01\mathbf{I},\:\forall l$.
For accuracy comparison, define the \textit{convergence error} as
the sum of the norms of the gradients over $\mathbf{F}$ and all $\mathbf{T}_{l}$'s. 

\begin{figure}
\centering{}%
\begin{minipage}[t]{0.48\textwidth}%
\includegraphics[width=80mm]{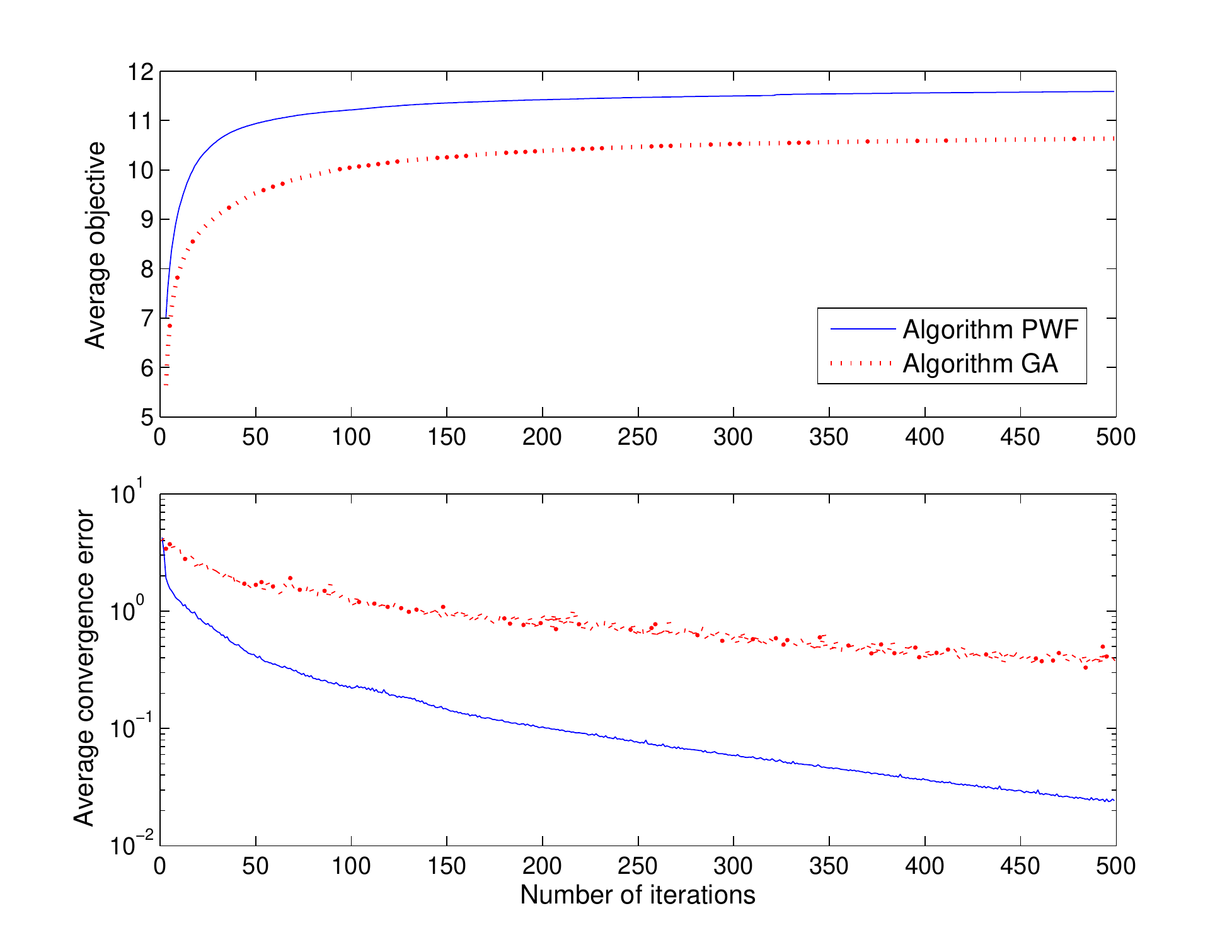}

\caption{\label{fig:twohop_BMAC_AF}Convergence comparison for the B-MAC AF
relay network in Fig. \ref{fig:B-MAC_Relay}}
\end{minipage}\hfill{}%
\begin{minipage}[t]{0.48\textwidth}%
\includegraphics[width=80mm]{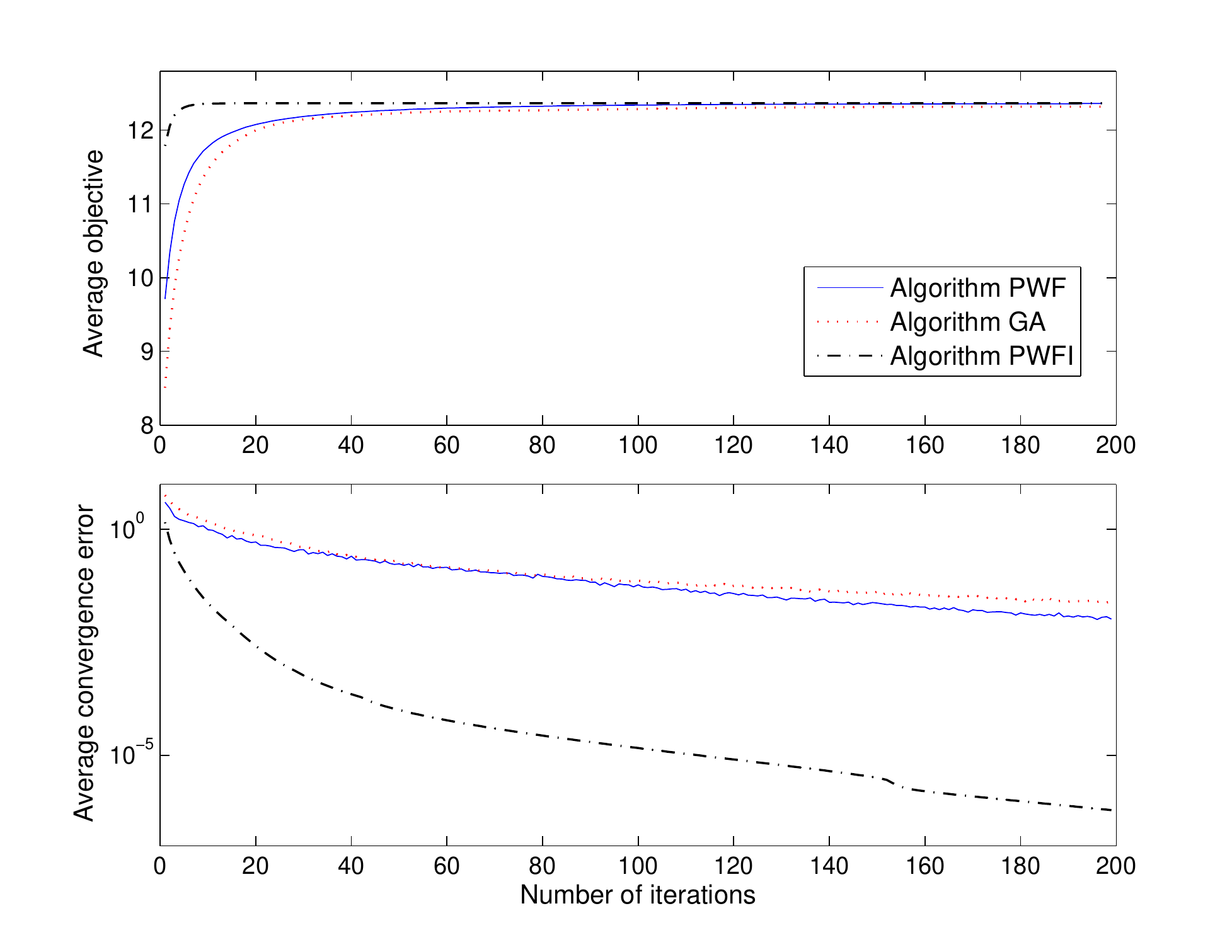}

\caption{\label{fig:twohop_MAC_AF}Convergence comparison for a two-hop MAC
AF relay network}
\end{minipage}
\end{figure}

We first demonstrate the advantages of Algorithm PWF over GA described
in Section \ref{sub:Baseline-Algorithm-(Gradient}. Consider the B-MAC
AF relay network in Fig. \ref{fig:B-MAC_Relay} with $Q=1$ (two-hop),
$n_{1}=1$ relay, and $L_{T_{l}}=4,\: L_{R_{l}}=2,\:\forall l$, $L_{1}=4$
antennas at each node. In Fig. \ref{fig:twohop_BMAC_AF}, we plot
the average objective value and the convergence error versus the number
of iterations. It can be observed that PWF has faster convergence
speed than GA. It also achieves a higher objective value. The complexity
order per iteration of PWF and GA is similar. However, the overall
complexity of PWF is lower because it avoids the linear search for
the gradient update of $\mathbf{\Sigma}$, and requires less iterations
to achieve the same accuracy.

In Fig. \ref{fig:twohop_MAC_AF}, we compare the convergence of Algorithm
GA, PWF and PWFI for a two-hop MAC relay network with $L=4$ sources,
$n_{1}=1$ relay and $L_{T_{l}}=2,\:\forall l$, $L_{1}=4$, $L_{R_{l}}=4,\:\forall l$
antennas at each node. The results show that PWFI has the fastest
convergence speed. PWFI also has the lowest complexity because it
avoids the linear search in the gradient update. 

\begin{figure}
\centering{}%
\begin{minipage}[t]{0.48\textwidth}%
\includegraphics[width=80mm,height=67mm]{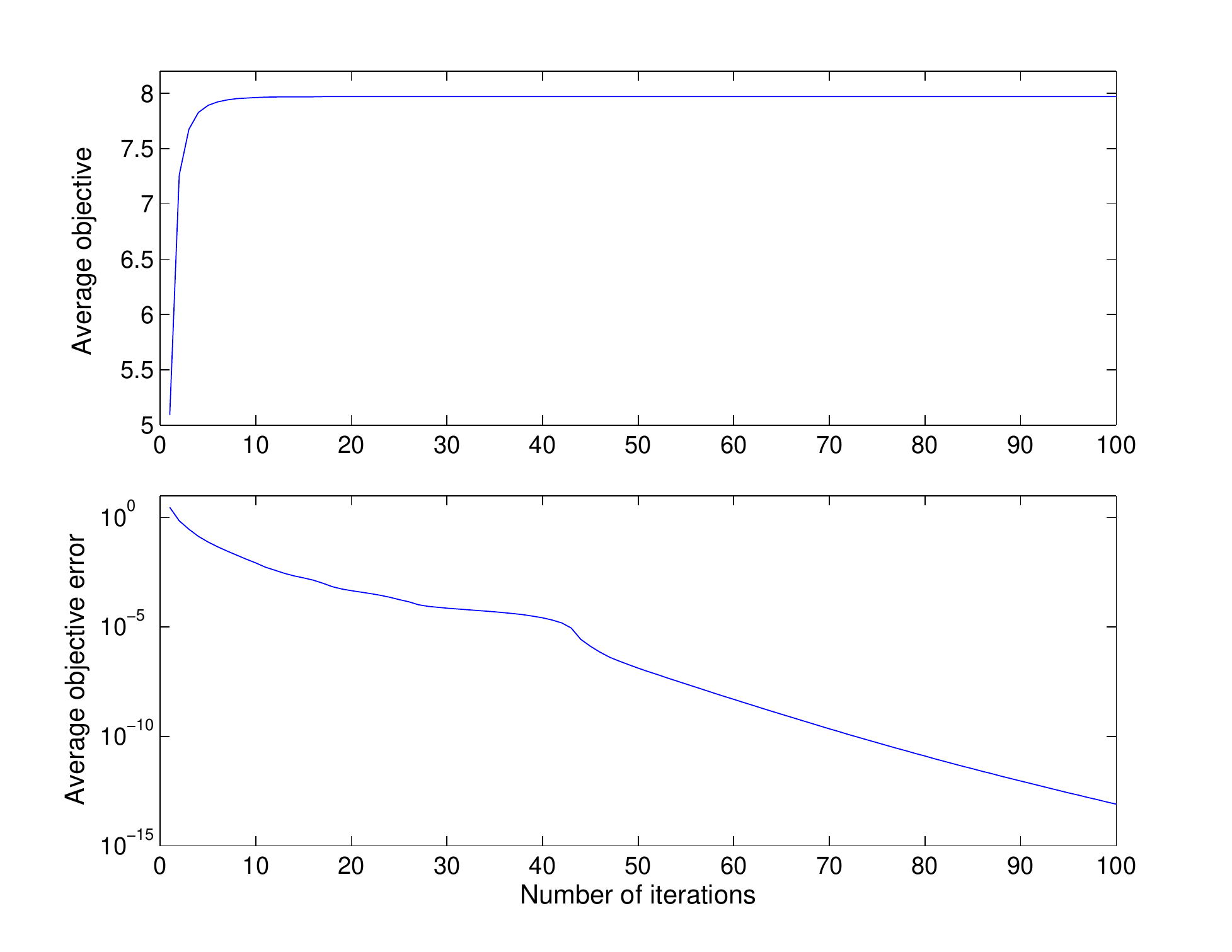}

\caption{\label{fig:threehop_BC_AF}Convergence of Algorithm PWF3 for a three-hop
BC AF relay network}
\end{minipage}\hfill{}%
\begin{minipage}[t]{0.48\textwidth}%
\includegraphics[width=80mm]{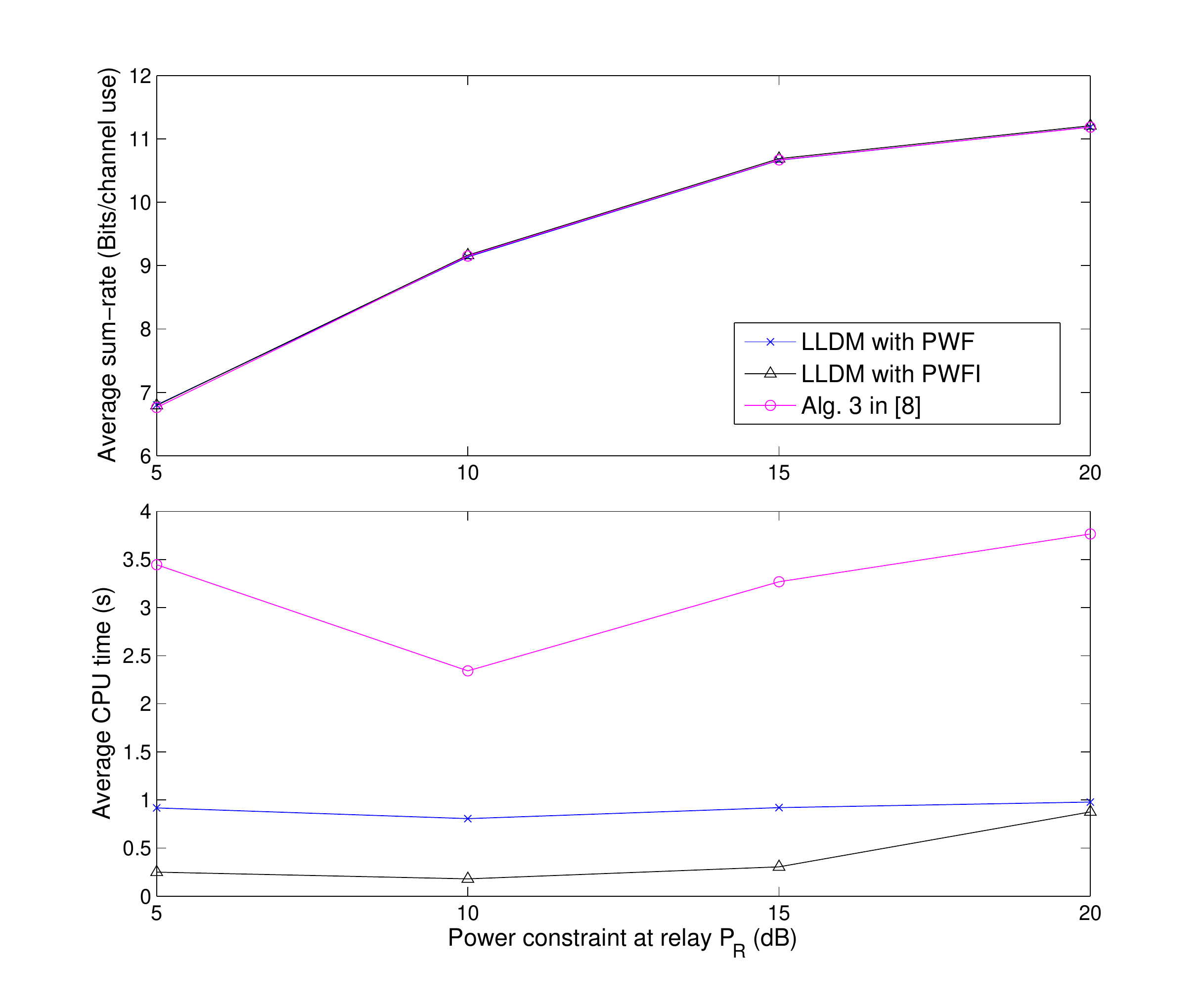}

\caption{\label{fig:twohop_BC_LLDM}Average sum-rate/CPU time versus power
constraint at relay for a two-hop BC AF relay network}
\end{minipage}
\end{figure}

Finally, we verify the convergence of Algorithm PWF3 for a three-hop
BC relay network with $n_{q}=1,q=1,2$ relay at each relay cluster,
$L=4$ destinations and $L_{T_{l}}=4,\:\forall l$, $L_{q}=4,\: q=1,2$,
$L_{R_{l}}=2,\:\forall l$ antennas at each node. Fig. \ref{fig:threehop_BC_AF}
plots the average objective value and \textit{objective error}, defined
as the gap (in logarithmic scale) from the stationary point that the
algorithm converges to, versus the number of iterations. Algorithm
PWF3 quickly converges to a stationary point of problem (\ref{eq:mainpro})
with high accuracy.

\subsection{Advantages of the Proposed Local LDM with Duality-based Primal Algorithm}

\begin{figure}
\centering{}%
\begin{minipage}[t]{0.48\textwidth}%
\includegraphics[width=80mm]{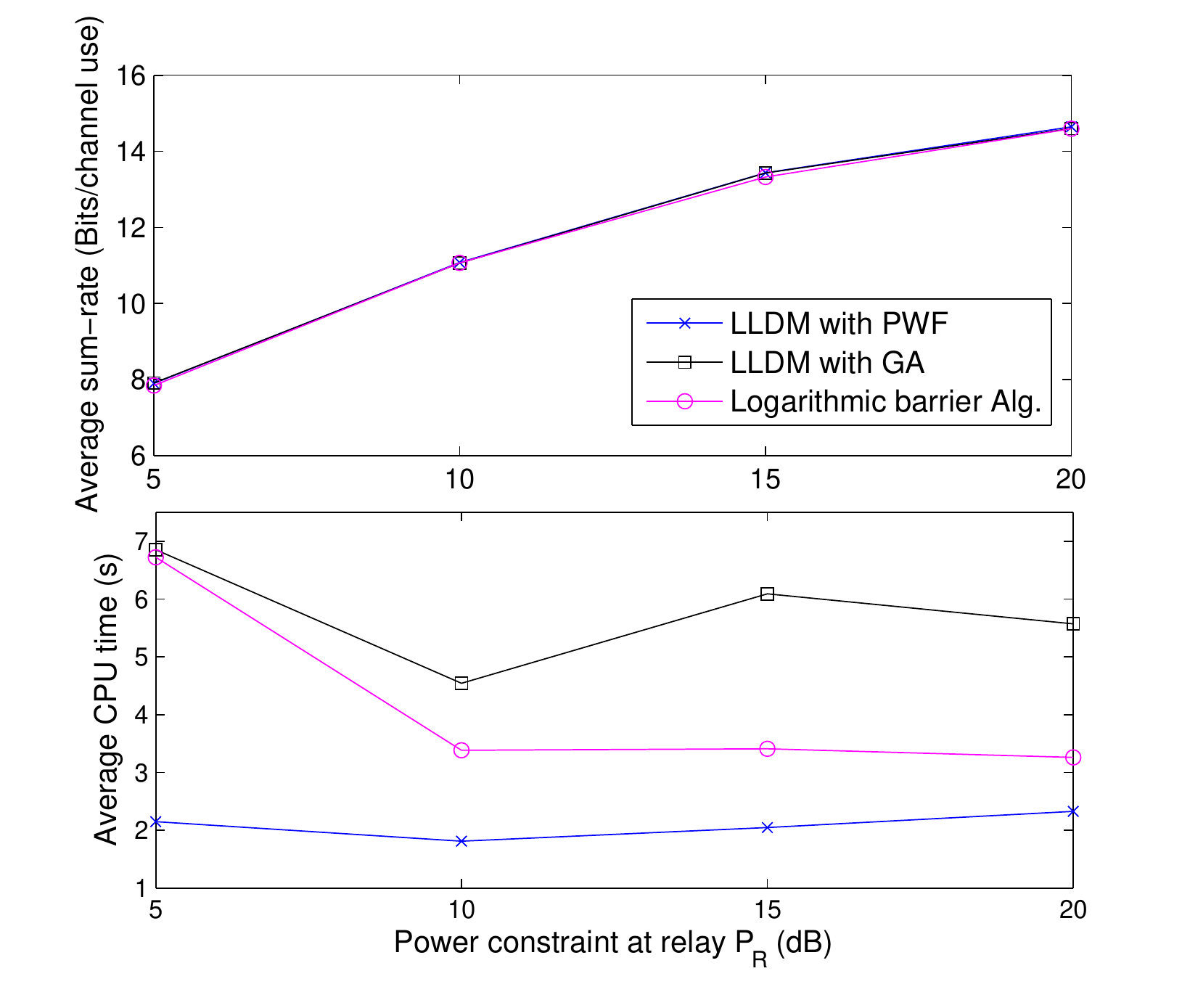}

\caption{\label{fig:twohop_BMAC_AF_LDM}Average sum-rate/CPU time vs. power
constraint at each relay for the B-MAC AF relay network in Fig. \ref{fig:B-MAC_Relay}}
\end{minipage}\hfill{}%
\begin{minipage}[t]{0.48\textwidth}%
\includegraphics[width=80mm]{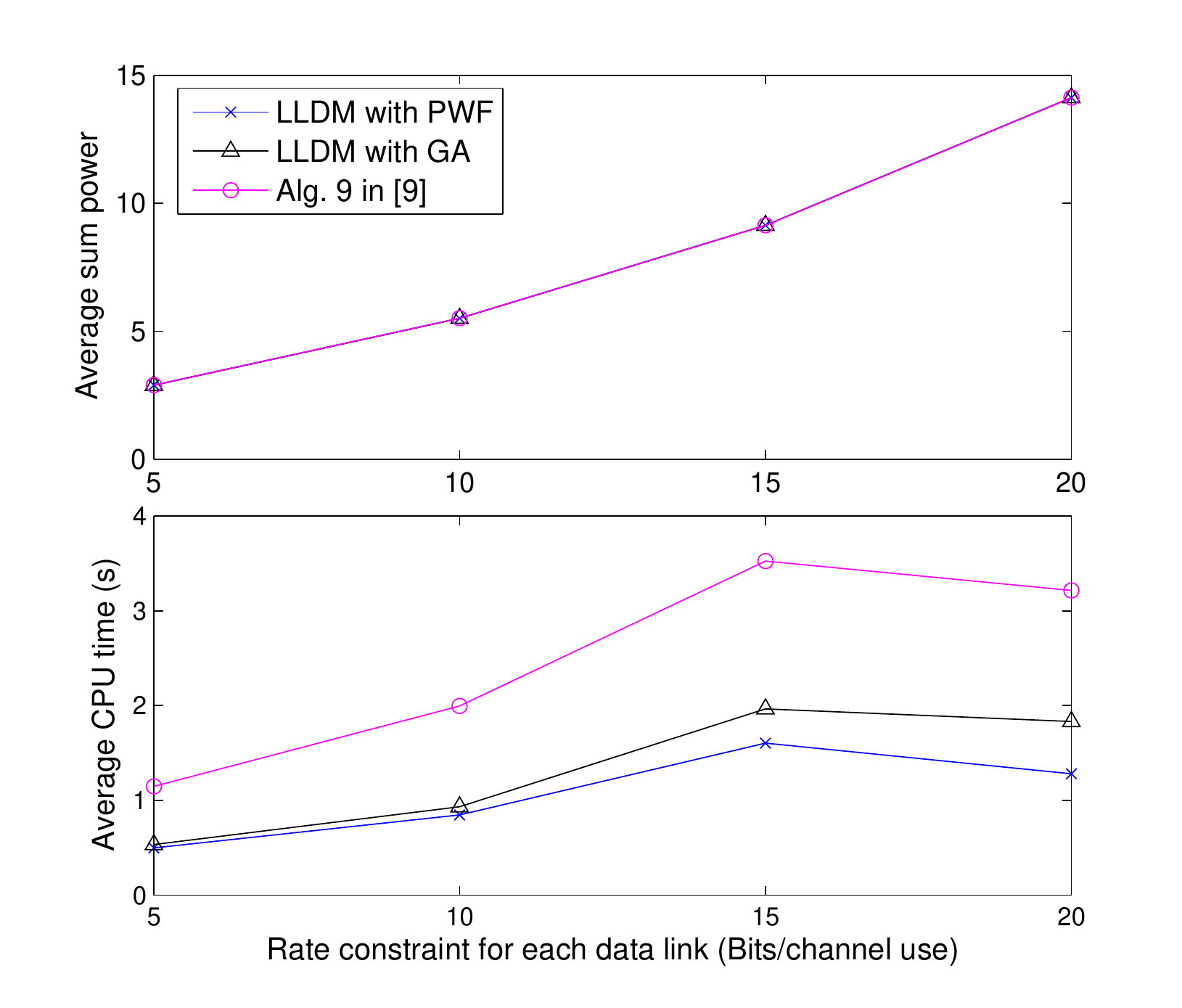}

\caption{\label{fig:twohop_MAC_AF_QoS}Average sum power/CPU time versus individual
rate constraint for a two-hop MAC AF relay network}
\end{minipage}
\end{figure}

In Fig. \ref{fig:threehop_BC_AF} and \ref{fig:twohop_BC_LLDM}, we
consider Problem (\ref{eq:P1}) with $\mu_{l}=1,\:\forall l$ (i.e.,
sum-rate maximization) and illustrate the advantages of the proposed
local LDM with duality-based primal algorithms (i.e., Algorithm PWF
or PWFI). The following baseline algorithms are compared.
\begin{itemize}
\item Baseline 1 (Algorithm 3 in \cite{Yu_TSP10_GWFRL}): Algorithm 3 in
\cite{Yu_TSP10_GWFRL} is based on logarithmic barrier method \cite{Boyd_04Book_Convex_optimization}
and is designed for sum-rate maximization under power constraints
in two-hop BC relay networks.
\item Baseline 2 (logarithmic barrier algorithm): This is an extension of
the Algorithm 3 in \cite{Yu_TSP10_GWFRL} to B-MAC AF relay networks
using the logarithmic barrier method \cite{Boyd_04Book_Convex_optimization}.
\item Baseline 3: Local LDM with GA as the primal algorithm.
\end{itemize}

We first evaluate the performance in a two-hop BC relay network with
$n_{1}=1$ relay, $L=4$ destinations and $L_{T_{l}}=4,\:\forall l$,
$L_{1}=4$, $L_{R_{l}}=2,\:\forall l$ antennas at each node. Assume
a sum power constraint at the source: $\sum_{l=1}^{4}\textrm{Tr}\left(\mathbf{\Sigma}_{l}\right)\leq10$,
and a power constraint at the relay: $\textrm{Tr}\left(\mathbf{\Sigma}_{3}\right)\leq P_{R}$.
In Fig. \ref{fig:twohop_BC_LLDM}, we plot the average sum-rate, and
the CPU time required to achieve the same accuracy, versus the power
constraint at relay $P_{R}$. All algorithms achieve similar sum-rate.
However, LLDM with both PWF and PWFI clearly outperform the Algorithm
3 in \cite{Yu_TSP10_GWFRL} in terms of CPU time.

In Fig. \ref{fig:twohop_BMAC_AF_LDM}, we evaluate the sum-rate and
CPU time performance in the B-MAC AF relay network in Fig. \ref{fig:twohop_BMAC_AF}
with $Q=1$ (two-hop), $n_{1}=2$ relays, $L_{1,j}=2,j=1,2$ antennas
at each relay, and $L_{T_{l}}=4,\: L_{R_{l}}=2,\:\forall l$ antennas
at each source/destination node. Consider individual power constraints
at each node: $\sum_{l=1}^{2}\textrm{Tr}\left(\mathbf{\Sigma}_{l}\right)\leq10$,
$\textrm{Tr}\left(\mathbf{\Sigma}_{3}\right)\leq10$, $\textrm{Tr}\left(\mathbf{\Sigma}_{1}^{R}\textrm{BlockDiag}\left[\mathbf{I}_{1},\mathbf{0}_{2}\right]\right)\leq P_{R}$
and $\textrm{Tr}\left(\mathbf{\Sigma}_{1}^{R}\textrm{BlockDiag}\left[\mathbf{0}_{1},\mathbf{I}_{2}\right]\right)\leq P_{R}$.
All algorithms achieve similar sum-rate. However, LLDM with PWF requires
the least CPU time.

In Fig. \ref{fig:twohop_MAC_AF_QoS}, we consider Problem (\ref{eq:P2})
with $\mathbf{\hat{W}}_{0}^{l}=\mathbf{I},\forall l,\:\mathbf{\hat{W}}_{q}=\mathbf{I},\forall q$
(i.e., sum power minimization), and $\mathcal{I}_{l}^{0}=\mathcal{I}^{0},\forall l$
for a two-hop MAC relay network with $L=3$ sources, $n_{1}=1$ relay
and $L_{T_{l}}=2,\:\forall l$, $L_{1}=4$, $L_{R_{l}}=4,\:\forall l$
antennas at each node. The proposed LLDM with PWF is compared to baseline
3 and baseline 4: Algorithm 9 in \cite{Yu_TSP10_GWFRL} (joint search
based on logarithmic barrier method). We plot the average sum power
and the CPU time versus the individual rate constraint ($\mathcal{I}^{0}$).
All algorithms achieve similar sum power. However, LLDM with PWF requires
the least CPU time.

\section{Conclusion\label{sec:Conlusion}}

We show that the achievable regions of a multi-hop MIMO B-MAC AF relay
network and its dual are the same under \textit{single network linear
constraint} or \textit{per-hop linear constraint}. Two dual transformations
are provided to calculate the dual input covariance and relay precoding
matrices. These results include the dualities in \cite{Jafar_TIT07_Dualrelay,Jafar_TCOM10_DualRelay,Rong_TWC11_dualrelay}
as special cases. Furthermore, our proof is simpler and reveals more
structural property of the duality. Based on the established duality
structure, we propose efficient algorithms for MIMO precoder optimization
in B-MAC AF relay networks. First, a unified optimization framework
is proposed based on the \textit{local Lagrange dual method} in \cite{Liu_11sTSP_MLC_localLDM}
so that we only need to focus on designing a primal algorithm to find
a stationary point of the unconstrained inner loop problem. Using
duality, we characterize the polite water-filling (PWF) structure
of the input covariance matrices at a stationary point of the inner
loop problem. Then, the duality and PWF are exploited to design efficient
primal algorithms. The proposed local LDM with duality-based primal
algorithms has lower computation cost and faster convergence speed
than the conventional step-size based iterative algorithms.

\appendix

\subsection{Proof of Theorem \ref{thm:Network-Equivalence} \label{sub:Proof-of-TheoremNE}}

It follows from $\mathbf{\hat{F}}_{q}=\mathbf{F}_{q}^{\dagger}$ that
$\mathbf{\hat{B}}_{q,q^{'}}=\mathbf{B}_{q,q^{'}},\:\forall q^{'}\geq q$.
Hence the equivalent channel of the dual network is $\hat{\mathbf{H}}_{l,k}=\mathbf{H}_{0}^{l\dagger}\mathbf{B}_{1,Q}^{\dagger}\mathbf{H}_{Q}^{k\dagger}=\mathbf{\check{H}}_{k,l}^{\dagger},\forall l,k$.
By (\ref{eq:whiteMG-1}), the covariance of the equivalent noise of
dual link $l$ is $\mathbf{\hat{W}}_{0}^{l}+\sum_{q=1}^{Q}\mathbf{H}_{0}^{l\dagger}\mathbf{\hat{B}}_{1,q}^{\dagger}\mathbf{\hat{W}}_{q}\mathbf{\hat{B}}_{1,q}\mathbf{H}_{0}^{l}=\mathbf{\hat{W}}_{0}^{l}+\sum_{q=1}^{Q}\mathbf{H}_{0}^{l\dagger}\mathbf{B}_{1,q}^{\dagger}\mathbf{\hat{W}}_{q}\mathbf{B}_{1,q}\mathbf{H}_{0}^{l}=\hat{\mathbf{W}}_{l}^{'}$.
Note that 
\begin{eqnarray*}
\sum_{q=1}^{Q}\textrm{Tr}\left(\mathbf{\mathbf{\hat{\Sigma}}}_{q}^{R}\mathbf{W}_{q}\right) & = & \sum_{q=1}^{Q}\sum_{l=1}^{L}\textrm{Tr}\left(\hat{\mathbf{B}}_{q,Q}^{\dagger}\mathbf{H}_{Q}^{l\dagger}\mathbf{\hat{\Sigma}}_{l}\mathbf{H}_{Q}^{l}\hat{\mathbf{B}}_{q,Q}\mathbf{W}_{q}\right)+\sum_{q=1}^{Q}\sum_{q^{'}=q}^{Q}\textrm{Tr}\left(\mathbf{\hat{B}}_{q,q^{'}}^{\dagger}\mathbf{\hat{W}}_{q^{'}}\mathbf{\hat{B}}_{q,q^{'}}\mathbf{W}_{q}\right)\\
 & = & \sum_{l=1}^{L}\textrm{Tr}\left(\mathbf{\hat{\Sigma}}_{l}\sum_{q=1}^{Q}\mathbf{H}_{Q}^{l}\mathbf{B}_{q,Q}\mathbf{W}_{q}\mathbf{B}_{q,Q}^{\dagger}\mathbf{H}_{Q}^{l\dagger}\right)+\sum_{q=1}^{Q}\sum_{q^{'}=q}^{Q}b_{q,q^{'}}.
\end{eqnarray*}
Hence, the linear constraint in (\ref{eq:DefdualSNLC}) can be expressed
as $\sum_{l=1}^{L}\textrm{Tr}\left(\mathbf{\hat{\Sigma}}_{l}\mathbf{W}_{l}^{'}\right)+P_{C}\leq P_{T}$.
The above proves the equivalence between the dual B-MAC AF relay network
under constraint (\ref{eq:DefdualSNLC}) and the B-MAC IFN in (\ref{eq:Dual_EquiBmac}).
The equivalence between the B-MAC AF relay network under constraint
(\ref{eq:DefSNLC}) and the B-MAC IFN in (\ref{eq:EquiBmac}) can
be proved similarly.

\subsection{\label{sub:Proof-of-Theorem_NDS}Proof of Theorem \ref{thm:Network-Dual-Scaling}}

Recall that $\mathbf{t}_{l,m}$, $\mathbf{r}_{l,m}$, $\mathbf{\Psi}$
and $\mathbf{D}$ are obtained using (\ref{eq:Decomsig}), (\ref{eq:MMSErev1G}),
(\ref{eq:faiG}) and (\ref{eq:DG}) in Definition \ref{def:The-covariance-transformation}
with parameters $\left\{ \left[\mathbf{\check{H}}_{l,k}\right],\left[\mathbf{W}_{l}^{'}\right],\left[\hat{\mathbf{W}}_{l}^{'}\right]\right\} $.
Since $\hat{\mathbf{\Sigma}}\left(\mathbf{d}\right)$ is the covariance
transformation of $\mathbf{\Sigma}$ obtained by Definition \ref{def:The-covariance-transformation}
with parameters $\left\{ \left[\mathbf{\check{H}}_{l,k}\right],\left[\mathbf{W}_{l}^{'}\right],\left[\hat{\mathbf{W}}_{l}^{'}\left(\mathbf{d}\right)\right]\right\} $,
using the fact that $\mathbf{t}_{l,m}$, $\mathbf{r}_{l,m}$, $\mathbf{\Psi}$
and $\mathbf{D}$ only depends on $\mathbf{\Sigma}$ and $\left\{ \left[\mathbf{\check{H}}_{l,k}\right],\left[\mathbf{W}_{l}^{'}\right]\right\} $,
we have 
\[
\hat{\mathbf{\Sigma}}_{l}\left(\mathbf{d}\right)=\sum_{m=1}^{M_{l}}\bar{q}_{l,m}\mathbf{r}_{l,m}\mathbf{r}_{l,m}^{\dagger},l=1,...,L,
\]
where $\left\{ \bar{q}_{l,m}\right\} $ is given by 
\begin{equation}
\mathbf{\bar{q}}=\left(\mathbf{D}^{-1}-\mathbf{\Psi}^{T}\right)^{-1}\left(\sum_{q=0}^{Q-1}d_{q+1}\mathbf{\hat{n}}_{q}+\mathbf{\hat{n}}_{Q}\right),\label{eq:qbar}
\end{equation}
and $\mathbf{\hat{n}}_{q}$'s are defined in (\ref{eq:Defnqhead}).
Note that 
\begin{eqnarray}
 &  & \sum_{l=1}^{L}\textrm{Tr}\left(\mathbf{\mathbf{\hat{\Sigma}}}_{l}\left(\mathbf{d}\right)\mathbf{W}_{Q+1}^{l}\right)=\sum_{l=1}^{L}\sum_{m=1}^{M_{l}}\mathbf{r}_{l,m}^{\dagger}\mathbf{W}_{Q+1}^{l}\mathbf{r}_{l,m}\bar{q}_{l,m}\nonumber \\
 & \overset{\textrm{a}}{=} & \mathbf{n}_{Q}^{T}\left(\mathbf{D}^{-1}-\mathbf{\Psi}^{T}\right)^{-1}\left(\sum_{q^{'}=0}^{Q-1}d_{q^{'}+1}\mathbf{\hat{n}}_{q^{'}}+\mathbf{\hat{n}}_{Q}\right)\nonumber \\
 & \overset{\textrm{b}}{=} & \sum_{q^{'}=0}^{Q-1}d_{q^{'}+1}a_{Q,q^{'}}+a_{Q,Q},\label{eq:Indscale1}
\end{eqnarray}
where (\ref{eq:Indscale1}-a) follows from (\ref{eq:Defnq}) and (\ref{eq:qbar}),
and (\ref{eq:Indscale1}-b) follows from the definition of $a_{q,q^{'}}$'s
in (\ref{eq:Defaq}). Hence the first linear constraint in (\ref{eq:IndLconsdualtx})
can be expressed as
\begin{equation}
\sum_{q^{'}=0}^{Q-1}d_{q^{'}+1}a_{Q,q^{'}}+a_{Q,Q}=P_{Q}^{\textrm{tx}}.\label{eq:Indscale1ina}
\end{equation}
Similarly, $\forall q=1,...,Q$, 
\begin{eqnarray}
 &  & \textrm{Tr}\left(\mathbf{\mathbf{\hat{\Sigma}}}_{q}^{R}\left(\mathbf{d}\right)\mathbf{W}_{q}\right)\nonumber \\
 & = & \sum_{l=1}^{L}\sum_{m=1}^{M_{l}}\mathbf{r}_{l,m}^{\dagger}\mathbf{H}_{Q}^{l}\mathbf{B}_{q,Q}\mathbf{W}_{q}\mathbf{B}_{q,Q}^{\dagger}\mathbf{H}_{Q}^{l\dagger}\mathbf{r}_{l,m}\bar{q}_{l,m}+\sum_{q^{'}=q}^{Q}d_{q^{'}+1}\textrm{Tr}\left(\mathbf{B}_{q,q^{'}}^{\dagger}\mathbf{\hat{W}}_{q^{'}}\mathbf{B}_{q,q^{'}}\mathbf{W}_{q}\right)\nonumber \\
 & = & \mathbf{n}_{q-1}^{T}\left(\mathbf{D}^{-1}-\mathbf{\Psi}^{T}\right)^{-1}\left(\sum_{q^{'}=0}^{Q-1}d_{q^{'}+1}\mathbf{\hat{n}}_{q^{'}}+\mathbf{\hat{n}}_{Q}\right)+\sum_{q^{'}=q}^{Q}d_{q^{'}+1}b_{q,q^{'}}\label{eq:Indscaleq}\\
 & = & \sum_{q^{'}=0}^{Q-1}d_{q^{'}+1}a_{q-1,q^{'}}+a_{q-1,Q}+\sum_{q^{'}=q}^{Q}d_{q^{'}+1}b_{q,q^{'}},\nonumber 
\end{eqnarray}
where (\ref{eq:Indscaleq}) follows from (\ref{eq:Defaq}) and (\ref{eq:Defbq}),
and thus the other linear constraints in (\ref{eq:IndLconsdualtx})
can be expressed as
\begin{equation}
\sum_{q^{'}=0}^{Q-1}d_{q^{'}+1}a_{q-1,q^{'}}+a_{q-1,Q}+\sum_{q^{'}=q}^{Q}d_{q^{'}+1}b_{q,q^{'}}=d_{q}P_{q-1}^{\textrm{tx}},\: q=1,...,Q.\label{eq:Indscaleqina}
\end{equation}

Now replace $P_{q}^{\textrm{tx}}$'s in (\ref{eq:Indscale1ina}) and
(\ref{eq:Indscaleqina}) with $\lambda P_{q}^{\textrm{tx}}$'s, we
obtain another set of equations
\begin{eqnarray}
\sum_{q^{'}=0}^{Q-1}d_{q^{'}+1}a_{Q,q^{'}}+a_{Q,Q} & = & \lambda P_{Q}^{\textrm{tx}},\label{eq:lamdEqu}\\
\sum_{q^{'}=0}^{Q-1}d_{q^{'}+1}a_{q-1,q^{'}}+a_{q-1,Q}+\sum_{q^{'}=q}^{Q}d_{q^{'}+1}b_{q,q^{'}} & = & d_{q}\lambda P_{q-1}^{\textrm{tx}},\: q=1,...,Q.\nonumber 
\end{eqnarray}
It can be verified that the equations in (\ref{eq:lamdEqu}) forms
the eigensystem $\mathbf{A}\mathbf{d}=\lambda\mathbf{d}$ in (\ref{eq:Keyeigensys}),
where $\mathbf{A}\in\mathbb{R}_{+}^{\left(Q+1\right)\times Q+1}$
is defined in (\ref{eq:DefA}), $\mathbf{d}=\left[d_{1},...,d_{Q},1\right]^{T}$.
Note that $a_{Q,0}>0$ due to nonsingularity of $\mathbf{W}_{Q+1}^{l}$'s
and $\mathbf{\hat{W}}_{0}^{l}$'s. Furthermore, since $\mathbf{W}_{q}$'s
and $\mathbf{\hat{W}}_{q}$'s are assumed to be non-singular and $\mathbf{F}_{q}\neq0,\: q=1,...,Q$,
we have $b_{q,q}>0,\: q=1,...,Q$. Using these facts, the following
lemma can be proved.
\begin{lem}
\label{lem:PropertyA}The following is true for $\mathbf{A}$ defined
in (\ref{eq:DefA}). 1) $\mathbf{A}\geq0$ and $\mathbf{A}\neq0$.
2) For any $\lambda$ and $\mathbf{d}$ satisfying $\mathbf{A}\mathbf{d}=\lambda\mathbf{d}$,
if any element of $\mathbf{d}$ is zero, then $\mathbf{d}=0$.
\end{lem}

It follows from Lemma \ref{lem:PropertyA} and the Perron\textendash{}Frobenius
theorem \cite[Chp. 8]{Meyer_2000_MatrixAna} that the maximum eigenvalue
$\lambda_{\textrm{max}}$ and the associated dominant eigenvector
$\mathbf{\tilde{d}}^{'}$ of $\mathbf{A}$ satisfies $\lambda_{\textrm{max}}>0$
and $\mathbf{\tilde{d}}^{'}>0$. Then we can always obtain a scaled
eigenvector $\mathbf{\tilde{d}}=\mathbf{\tilde{d}}^{'}/\tilde{d}_{Q+1}^{'}=\left[\tilde{d}_{1},...,\tilde{d}_{Q},1\right]^{T}$
that satisfies $\mathbf{A}\mathbf{\tilde{d}}=\lambda_{\textrm{max}}\mathbf{\tilde{d}}$.
On the other hand, it follows from (\ref{eq:SNLCfixd}) that $\lambda_{\textrm{max}}$
is equal to 1, which indicates that $\hat{\mathbf{\Sigma}}\left(\tilde{\mathbf{d}}\right)$
and $\mathbf{\bar{\hat{F}}}$ satisfies (\ref{eq:IndLconsdualtx}).

\subsection{Proof of Theorem \ref{thm:dualnetPro} \label{sub:Proof-of-Theoremdualpro}}

By Corollary \ref{cor:PWF_MLC}, $\mathbf{\bar{\Sigma}}$ satisfies
the PWF structure for fixed $\bar{\mathbf{F}}$ and the KKT condition
in (\ref{eq:KKTsigma}) with $\mathbf{F}=\bar{\mathbf{F}}$. To prove
that $\mathbf{\bar{\Sigma}},\bar{\mathbf{F}}$ is a stationary point,
we only need to further show that $\mathbf{\bar{\Sigma}},\bar{\mathbf{F}}$
satisfies 
\begin{equation}
\nabla_{\mathbf{F}_{q}\left(j\right)}L_{\lambda}\left(\mathbf{\bar{\Sigma}},\bar{\mathbf{F}}\right)=0,\:\forall q=1,...,Q,j=1,...,n_{q}.\label{eq:KKTF}
\end{equation}
Since $\mathbf{\mathbf{\bar{\hat{\Sigma}}}},\mathbf{\bar{\hat{F}}}$
is a stationary point of the dual problem (\ref{eq:mainpro_dual}),
we have $\nabla_{\hat{\mathbf{F}}_{q}\left(j\right)}\hat{L}_{\lambda}\left(\mathbf{\mathbf{\bar{\hat{\Sigma}}}},\mathbf{\bar{\hat{F}}}\right)=0,\forall q,j$.
In the following, we prove that $\nabla_{\mathbf{F}_{q}}L_{\lambda}\left(\mathbf{\bar{\Sigma}},\bar{\mathbf{F}}\right)=\left(\nabla_{\hat{\mathbf{F}}_{q}}\hat{L}_{\lambda}\left(\mathbf{\mathbf{\bar{\hat{\Sigma}}}},\mathbf{\bar{\hat{F}}}\right)\right)^{*},\forall q$,
from which (\ref{eq:KKTF}) follows immediately.

Note that for fixed $\bar{\mathbf{F}}_{q^{'}},\forall q^{'}\neq q$
and $\mathbf{\bar{\hat{F}}}_{q^{'}},\forall q^{'}\neq q$, problem
(\ref{eq:mainpro}) and (\ref{eq:mainpro_dual}) can be equivalent
to the inner loop problem and its dual for a two-hop B-MAC AF relay
network. Hence, without loss of generality, we only need to consider
the two-hop case (i.e., $Q=1$) and prove that $\nabla_{\mathbf{F}}L_{\lambda}\left(\mathbf{\bar{\Sigma}},\bar{\mathbf{F}}\right)=\left(\nabla_{\hat{\mathbf{F}}}\hat{L}_{\lambda}\left(\mathbf{\mathbf{\bar{\hat{\Sigma}}}},\mathbf{\bar{\hat{F}}}\right)\right)^{*}$.
Using (\ref{eq:sigmhead1}) in Corollary \ref{cor:PWF_MLC} and the
expression of $\nabla_{\mathbf{F}}L_{\lambda}\left(\mathbf{\bar{\Sigma}},\bar{\mathbf{F}}\right)$
in (\ref{eq:GF}), it can be shown that
\begin{eqnarray}
\nabla_{\mathbf{F}}L_{\lambda}\left(\mathbf{\bar{\Sigma}},\bar{\mathbf{F}}\right) & = & -2\sum_{l=1}^{L}\sum_{k=1}^{L}\phi_{l,k}\bar{\hat{\mathbf{S}}}_{l}\bar{\mathbf{F}}\bar{\mathbf{S}}_{k}-2\sum_{l=1}^{L}\bar{\hat{\mathbf{S}}}_{l}\bar{\mathbf{F}}\mathbf{W}_{1}-2\sum_{l=1}^{L}\bar{\hat{\mathbf{S}}}_{l}\bar{\mathbf{F}}\bar{\mathbf{S}}_{l}\nonumber \\
 &  & +2\sum_{l=1}^{L}w_{l}\mathbf{H}_{1}^{l\dagger}\bar{\mathbf{\Omega}}_{l}^{-1}\mathbf{H}_{1}^{l}\bar{\mathbf{F}}\mathbf{H}_{0}^{l}\mathbf{\Sigma}_{l}\mathbf{H}_{0}^{l\dagger}-2\sum_{l=1}^{L}\hat{\mathbf{W}}_{1}\bar{\mathbf{F}}\bar{\mathbf{S}}_{l},\label{eq:GF1}
\end{eqnarray}
where $\bar{\mathbf{S}}_{k}=\mathbf{H}_{0}^{k}\mathbf{\Sigma}_{k}\mathbf{H}_{0}^{k\dagger}$
and $\bar{\hat{\mathbf{S}}}_{l}=\mathbf{H}_{1}^{l\dagger}\mathbf{\mathbf{\bar{\hat{\Sigma}}}}_{l}\mathbf{H}_{1}^{l}$.
Similar, 
\begin{eqnarray}
\nabla_{\hat{\mathbf{F}}}\hat{L}_{\lambda}\left(\mathbf{\mathbf{\bar{\hat{\Sigma}}}},\mathbf{\bar{\hat{F}}}\right) & = & -2\sum_{l=1}^{L}\sum_{k=1}^{L}\phi_{k,l}\bar{\mathbf{S}}_{l}\mathbf{\bar{\hat{F}}}\bar{\hat{\mathbf{S}}}_{k}-2\sum_{l=1}^{L}\bar{\mathbf{S}}_{l}\mathbf{\bar{\hat{F}}}\hat{\mathbf{W}}_{1}-2\sum_{l=1}^{L}\bar{\mathbf{S}}_{l}\mathbf{\bar{\hat{F}}}\bar{\hat{\mathbf{S}}}_{l}\nonumber \\
 &  & +2\sum_{l=1}^{L}w_{l}\mathbf{H}_{0}^{l}\mathbf{\bar{\hat{\Omega}}}_{l}^{-1}\mathbf{H}_{0}^{l\dagger}\mathbf{\bar{\hat{F}}}\mathbf{H}_{1}^{l\dagger}\mathbf{\mathbf{\bar{\hat{\Sigma}}}}_{l}\mathbf{H}_{1}^{l}-2\sum_{l=1}^{L}\mathbf{W}_{1}\mathbf{\bar{\hat{F}}}\bar{\hat{\mathbf{S}}}_{l}.\label{eq:GFh1}
\end{eqnarray}
It follows from (\ref{eq:PWF_MGL}) and (\ref{eq:PWF_MLC_dual}) in
Corollary \ref{cor:PWF_MLC} that
\begin{equation}
\mathbf{H}_{1}^{l\dagger}\bar{\mathbf{\Omega}}_{l}^{-1}\mathbf{H}_{1}^{l}\bar{\mathbf{F}}\mathbf{H}_{0}^{l}\mathbf{\Sigma}_{l}\mathbf{H}_{0}^{l\dagger}=\mathbf{H}_{1}^{l\dagger}\mathbf{\mathbf{\bar{\hat{\Sigma}}}}_{l}\mathbf{H}_{1}^{l}\bar{\mathbf{F}}\mathbf{H}_{0}^{l}\mathbf{\bar{\hat{\Omega}}}_{l}^{-1}\mathbf{H}_{0}^{l\dagger}.\label{eq:KeyEqGFequ}
\end{equation}
Combining (\ref{eq:GF1}-\ref{eq:KeyEqGFequ}) and $\mathbf{\bar{\hat{F}}}=\bar{\mathbf{F}}^{\dagger}$,
we have $\nabla_{\mathbf{F}}L_{\lambda}\left(\mathbf{\bar{\Sigma}},\bar{\mathbf{F}}\right)=\left(\nabla_{\hat{\mathbf{F}}}\hat{L}_{\lambda}\left(\mathbf{\mathbf{\bar{\hat{\Sigma}}}},\mathbf{\bar{\hat{F}}}\right)\right)^{*}$.
This completes the proof.


\begin{thebibliography}{10}
\providecommand{\url}[1]{#1}
\csname url@samestyle\endcsname
\providecommand{\newblock}{\relax}
\providecommand{\bibinfo}[2]{#2}
\providecommand{\BIBentrySTDinterwordspacing}{\spaceskip=0pt\relax}
\providecommand{\BIBentryALTinterwordstretchfactor}{4}
\providecommand{\BIBentryALTinterwordspacing}{\spaceskip=\fontdimen2\font plus
\BIBentryALTinterwordstretchfactor\fontdimen3\font minus
  \fontdimen4\font\relax}
\providecommand{\BIBforeignlanguage}[2]{{%
\expandafter\ifx\csname l@#1\endcsname\relax
\typeout{** WARNING: IEEEtran.bst: No hyphenation pattern has been}%
\typeout{** loaded for the language `#1'. Using the pattern for}%
\typeout{** the default language instead.}%
\else
\language=\csname l@#1\endcsname
\fi
#2}}
\providecommand{\BIBdecl}{\relax}
\BIBdecl

\bibitem{Sanguinetti_JSAC2012_AFsurvey}
L.~Sanguinetti, A.~D'Amico, and Y.~Rong, ``A tutorial on the optimization of
  amplify-and-forward {MIMO} relay systems,'' \emph{IEEE J. Select. Areas
  Commun.}, vol.~30, no.~8, pp. 1331--1346, Sep. 2012.

\bibitem{Hua_TWC2007_optAF}
X.~Tang and Y.~Hua, ``Optimal design of non-regenerative {MIMO} wireless
  relays,'' \emph{IEEE Trans. Wireless Commun.}, vol.~6, no.~4, pp. 1398--1407,
  Apr. 2007.

\bibitem{Yue_TWC10_NLrelay_DFE}
Y.~Rong, ``Optimal linear non-regenerative multi-hop {MIMO} relays with
  {MMSE-DFE} receiver at the destination,'' \emph{IEEE Tans. Wireless Commun.},
  vol.~9, no.~7, pp. 2268 --2279, Jul. 2010.

\bibitem{Rong_TSP11_AFQos}
------, ``Multihop nonregenerative {MIMO} relays - {QoS} considerations,''
  \emph{IEEE Trans. Signal Processing}, vol.~59, no.~1, pp. 290--303, Jan.
  2011.

\bibitem{Sanguinetti_TSP12_optAF}
L.~Sanguinetti and A.~D'Amico, ``Power allocation in two-hop
  amplify-and-forward {MIMO} relay systems with {QoS} requirements,''
  \emph{IEEE Trans. Signal Processing}, vol.~60, no.~5, pp. 2494--2507, May
  2012.

\bibitem{Zhang_TOC11_optAFOFDM}
W.~Zhang, U.~Mitra, and M.~Chiang, ``Optimization of amplify-and-forward
  multicarrier two-hop transmission,'' \emph{IEEE Trans. Commun.}, vol.~59,
  no.~5, pp. 1434--1445, May 2011.

\bibitem{Fu_TSP11_AFmultiRelays}
Y.~Fu, L.~Yang, W.-P. Zhu, and C.~Liu, ``Optimum linear design of two-hop
  {MIMO} relay networks with {QoS} requirements,'' \emph{IEEE Trans. Signal
  Processing}, vol.~59, no.~5, pp. 2257--2269, May 2011.

\bibitem{Phan_TWC09_PowAFMAC}
K.~Phan, T.~Le-Ngoc, S.~Vorobyov, and C.~Tellambura, ``Power allocation in
  wireless multi-user relay networks,'' \emph{IEEE Trans. Wireless Commun.},
  vol.~8, no.~5, pp. 2535--2545, May 2009.

\bibitem{Yu_TSP10_GWFRL}
Y.~Yu and Y.~Hua, ``Power allocation for a {MIMO} relay system with
  multiple-antenna users,'' \emph{IEEE Trans. Signal Processing}, vol.~58, pp.
  2823 -- 2835, May 2010.

\bibitem{Boyd_04Book_Convex_optimization}
S.~Boyd and L.~Vandenberghe, \emph{Convex Optimization}.\hskip 1em plus 0.5em
  minus 0.4em\relax Cambridge University Press, 2004.

\bibitem{Jafar_TIT07_Dualrelay}
S.~A. Jafar, K.~S. Gomadam, and C.~Huang, ``Duality and rate optimization for
  multiple access and broadcast channels with amplify-and-forward relays,''
  \emph{IEEE Trans. Inf. Theory,}, vol.~53, no.~10, pp. 3350 --3370, oct. 2007.

\bibitem{Jafar_TCOM10_DualRelay}
K.~Gomadam and S.~Jafar, ``Duality of {MIMO} multiple access channel and
  broadcast channel with amplify-and-forward relays,'' \emph{IEEE Trans.
  Commun.,}, vol.~58, no.~1, pp. 211 --217, january 2010.

\bibitem{Rong_TWC11_dualrelay}
Y.~Rong and M.~Khandaker, ``On uplink-downlink duality of multi-hop {MIMO}
  relay channel,'' \emph{IEEE Trans. Wireless Commun.}, vol.~10, no.~6, pp.
  1923 --1931, june 2011.

\bibitem{Jafar_TIT09_NoissCorrelationAF}
K.~Gomadam and S.~Jafar, ``The effect of noise correlation in
  amplify-and-forward relay networks,'' \emph{IEEE Transactions on Information
  Theory,}, vol.~55, no.~2, pp. 731 --745, feb. 2009.

\bibitem{Liu_IT10s_Duality_BMAC}
\BIBentryALTinterwordspacing
A.~Liu, Y.~Liu, H.~Xiang, and W.~Luo, ``Duality, polite water-filling, and
  optimization for {MIMO B-MAC} interference networks and {iTree} networks,''
  \emph{submitted to IEEE Trans. Info. Theory}, Apr. 2010; revised Oct. 2010.
  [Online]. Available: \url{http://arxiv.org/abs/1004.2484}
\BIBentrySTDinterwordspacing

\bibitem{Liu_10sTSP_MLC}
------, ``Polite water-filling for weighted sum-rate maximization in {B-MAC}
  networks under multiple linear constraints,'' \emph{IEEE Trans. Signal
  Processing}, vol.~60, no.~2, pp. 834 --847, Feb. 2012.

\bibitem{Bertsekas_book99_NProgramming}
D.~P. Bertsekas, \emph{Nonlinear Programming}, 2nd~ed.\hskip 1em plus 0.5em
  minus 0.4em\relax Belmont, MA: Athena Scientific, 1999.

\bibitem{Hua_TSP11_AFchannelE}
T.~Kong and Y.~Hua, ``Optimal design of source and relay pilots for {MIMO}
  relay channel estimation,'' \emph{IEEE Trans. Signal Processing}, vol.~59,
  no.~9, pp. 4438--4446, Sept. 2011.

\bibitem{Liu_10sTSP_Fairness_rate_polit_WF}
A.~Liu, Y.~Liu, H.~Xiang, and W.~Luo, ``{MIMO B-MAC} interference network
  optimization under rate constraints by polite water-filling and duality,''
  \emph{IEEE Trans. Signal Processing}, vol.~59, no.~1, pp. 263 --276, Jan.
  2011.

\bibitem{Costa_IT83_Dirty_paper}
M.~Costa, ``Writing on dirty paper (corresp.),'' \emph{IEEE Trans. Info.
  Theory}, vol.~29, no.~3, pp. 439--441, 1983.

\bibitem{Telatar_EuroTrans_1999_MIMOCapacity}
E.~Telatar, ``Capacity of multi-antenna gaussian channels,'' \emph{Europ.
  Trans. Telecommu.}, vol.~10, pp. 585--595, Nov./Dec. 1999.

\bibitem{Rao_TOC07_netduality}
B.~Song, R.~Cruz, and B.~Rao, ``Network duality for multiuser {MIMO}
  beamforming networks and applications,'' \emph{IEEE Trans. Commun.}, vol.~55,
  no.~3, pp. 618--630, March 2007.

\bibitem{Liu_11sTSP_MLC_localLDM}
\BIBentryALTinterwordspacing
A.~Liu, V.~K.~N. Lau, and Y.~Liu, ``Local dual method for optimization of
  parallel {MIMO B-MAC} interference networks under multiple linear
  constraints,'' \emph{submitted to IEEE Trans. Signal Processing}, Sep. 2011;
  revised Apr. 2012. [Online]. Available:
  \url{http://www.ee.ust.hk/~eeknlau/HKUST-Office-HomePage/Publications.html}
\BIBentrySTDinterwordspacing

\bibitem{Luenberger_Book08_LNLProgramming}
D.~G. Luenberger and Y.~Ye, \emph{Linear and Nonlinear Programming},
  3rd~ed.\hskip 1em plus 0.5em minus 0.4em\relax New York: Springer, 2008.

\bibitem{Hjorungnes_TSP07_ComplexDiff}
A.~Hjorungnes and D.~Gesbert, ``Complex-valued matrix differentiation:
  Techniques and key results,'' \emph{IEEE Transactions on Signal Processing},
  vol.~55, no.~6, pp. 2740--2746, June 2007.

\bibitem{Meyer_2000_MatrixAna}
C.~D. Meyer, Ed., \emph{Matrix analysis and applied linear algebra}.\hskip 1em
  plus 0.5em minus 0.4em\relax Philadelphia, PA, USA: Society for Industrial
  and Applied Mathematics, 2000.

\end{thebibliography}
\end{document}